\documentclass[aps, prd, twocolumn, amsmath, amssymb, superscriptaddress,longbibliography,floatfix]{revtex4-2}

\usepackage[dvipsnames]{xcolor}
\usepackage{natbib}
\usepackage{graphicx}
\usepackage{enumitem}
\usepackage{siunitx}
\usepackage{booktabs}
\usepackage{multirow}
\usepackage{mathtools}
\usepackage{slashed} 
\usepackage{simplewick} 
\usepackage{simpler-wick} 
\usepackage[switch]{lineno}
\usepackage[utf8]{inputenc}
\usepackage{comment}

\definecolor{deepgreen}{rgb}{0.2,0.8,0.2}

\definecolor{deepblue}{rgb}{0.2,0.2,0.8}

\definecolor{deepred}{rgb}{0.8,0.2,0.2}

\newcommand{\Sec}[1]{Sec.~\ref{#1}}
\newcommand{\Secs}[1]{Secs.~\ref{#1}}
\newcommand{\App}[1]{App.~\ref{#1}}
\newcommand{\Apps}[1]{Apps.~\ref{#1}}
\newcommand{\Eq}[1]{Eq.~\ref{#1}}
\newcommand{\Eqs}[1]{Eqs.~\ref{#1}}
\newcommand{\Fig}[1]{Fig.~\ref{#1}}
\newcommand{\Figs}[1]{Figs.~\ref{#1}}

\newcommand{\vect}[1]{\boldsymbol{\mathbf{#1}}}
\newcommand{\half}{\frac{1}{2}}
\newcommand{\deltathree}{\delta^{(3)}}
\def\deltabar{{\mathchar '26\mkern -10mu\delta}}

\newcommand{\ddbar}{{\rm d}\hspace*{-0.15em}\bar{}\hspace*{0.1em}}
\newcommand{\deltabarthree}{\deltabar^{(3)}}

\newcommand{\dd}{{\mathrm{d}}}
\newcommand{\cc}{\text{c.c.}}

\usepackage{tikz} 
\usetikzlibrary{shapes,arrows,positioning,automata,backgrounds,calc,er,patterns}
\usepackage{tikz-feynman}
\tikzfeynmanset{compat=1.1.0}

\newcommand\myshade{30}
\colorlet{mylinkcolor}{red}
\colorlet{mycitecolor}{orange}
\colorlet{myurlcolor}{orange}

\usepackage[%
  bookmarks=true,
  colorlinks,
  linkcolor=mylinkcolor!\myshade!black,
  urlcolor=myurlcolor!\myshade!black,
  citecolor=mycitecolor!\myshade!black,
  plainpages=false,
  pdfpagelabels,
  final,
  breaklinks=true
]{hyperref}

\hypersetup{
pdftitle={Wake Forces},
pdfauthor={Ken Van Tilburg},
pdfkeywords={short-range forces, wake forces, dark matter, neutrinos}
}

\begin{document}
\title[Wake Forces]{Wake forces in a background of quadratically coupled mediators}

\author{Ken Van Tilburg}
\email{kenvt@nyu.edu, kvantilburg@flatironinstitute.org}
\affiliation{Center for Cosmology and Particle Physics, Department of Physics, New York University,
New York, NY 10003, USA}
\affiliation{Center for Computational Astrophysics, Flatiron Institute, New York, NY 10010, USA}

\date{\today}

\begin{abstract}
    Two particles can exert forces on each other when embedded in a sea of weakly-coupled particles. These ``wake forces'' occur whenever the source and target particles have quadratic interactions with the mediating particles; they are proportional to the ambient energy density, and typically have a range of order the characteristic de Broglie wavelength of the background. The effect can be understood as source particles causing a disturbance in the background waves---a wake---which subsequently interacts with the target particles. Wake forces can be mediated by bosons or fermions, can have spin dependence, may be attractive or repulsive, and have a generally anisotropic spatial profile and range that depends on the phase-space distribution of the ambient particles. In this work, I investigate the application of wake forces to dark matter searches, recast existing limits on short-range forces into leading constraints on dark matter with quadratic couplings, and sketch out potential experimental modifications to optimize sensitivity. Wake forces occur in the Standard Model: the presence of the cosmic neutrino background induces a millimeter-range force about 22 orders of magnitude weaker than gravity. Wake forces may also be relevant in condensed-matter and atomic physics.
\end{abstract}

\maketitle

\section{Introduction} \label{sec:intro}

The birth of the modern scientific method can arguably be traced to Galileo's gravitational experiments with inclined planes. His determination that the free-fall acceleration of objects is independent of their mass and composition, now known as the (weak) equivalence principle, is a cornerstone of the theory of gravitational interactions. Since then, swaths of experiments have been performed to test the equivalence principle, and the inverse-square-radius scaling of the two long-range forces of nature, gravity and electromagnetism. 

These efforts were further boosted by the realization that motivated theories beyond the Standard Model (SM) of particle physics---the QCD axion~\cite{Peccei:1977ur,Weinberg:1977ma,Wilczek:1977pj,Kim:1979if,Shifman:1979if,Dine:1981rt,Zhitnitsky:1980tq} and large extra dimensions~\cite{Arkani-Hamed:1998jmv,Antoniadis:1998ig} in particular---can exhibit small deviations from these predictions or even qualitatively new forces~\cite{Moody:1984ba}. These forces are typically mediated by low-mass particles---dilatons, moduli, axions, dark photons, or gravitons---which couple linearly but weakly to regular matter. Single-particle-exchange forces generally have a range equal to the Compton wavelength of the mediator; any such particle with gravitational-strength, spin-independent couplings to matter must be heavier than $5.11\,\mathrm{meV} \approx (38.6 \, \mathrm{\mu m})^{-1}$ from tests of the gravitational inverse-square law~\cite{Lee:2020zjt}.

These light, weakly-coupled fields can also make up the dark matter (DM), as they are naturally long-lived and have generic production mechanisms in the early universe~\cite{Dine:1982ah,Preskill:1982cy,Abbott:1982af}. Linearly-coupled DM fields generally give rise to temporally oscillating phenomena at a frequency equal to the mass. Light scalar DM causes oscillations of fundamental ``constants''~\cite{Arvanitaki:2014faa}, which in turn lead to time-varying energies~\cite{Arvanitaki:2014faa,Arvanitaki:2016fyj}, length scales~\cite{Arvanitaki:2015iga}, and forces~\cite{Graham:2015ifn,Arvanitaki:2014faa}, the latter through gradients in the field. 
Pseudoscalar DM can excite electromagnetic fields~\cite{Sikivie:1983ip}, spin resonances~\cite{Graham:2013gfa}, and acoustic modes~\cite{Arvanitaki:2021wjk}. At higher DM masses, current and near-future single-quantum detection techniques are sufficiently sensitive to search for DM absorption~\cite{Hochberg:2016sqx,Arvanitaki:2017nhi} and conversion~\cite{Baryakhtar:2018doz,Chiles:2021gxk}.

\emph{Quadratic} interactions between DM and SM operators $\mathcal{O}_\mathrm{SM}$ are also possible, and are the leading interactions if the linear coupling is forbidden by a symmetry of the DM field. Examples include a $\mathbb{Z}_2$ parity symmetry $\phi \mapsto - \phi$ for a real scalar field, forbidding $\phi \mathcal{O}_\mathrm{SM}$ but allowing $\phi^2 \mathcal{O}_\mathrm{SM}$. Likewise, a $\mathrm{U}(1)$ symmetry $\Phi \mapsto e^{i\alpha} \Phi$ of a complex scalar field implies a leading interaction of $|\Phi|^2 \mathcal{O}_\mathrm{SM}$ in the effective field theory (EFT). Fermions beyond the SM (other than sterile neutrinos) necessarily have quadratic couplings to SM operators because of fermion number symmetry. Notably, the QCD axion $a$ has an irreducible quadratic coupling to nucleons $N = (p,n)$ of the form $\mathcal{L} \supset (\widetilde{\sigma}/2) (a/f_a)^2 \bar{N} N$, with $\widetilde{\sigma} \approx 15 \, \mathrm{MeV}$ and $f_a$ the axion decay constant~\cite{Okawa:2021fto}. Finally, SM neutrinos couple quadratically to matter at low energies through the four-Fermi interaction:
\begin{align}
\hspace{-0.4em} \mathcal{L} &\supset -\frac{G_F}{\sqrt{2}} \sum_\psi \bar{\psi}\left(g^\text{V}_\psi \gamma^\mu - g^\text{A}_\psi \gamma^\mu \gamma^5 \right)\psi \, \bar{\nu}_i \gamma_\mu \left(1-\gamma^5\right) \nu_i \label{eq:L_nu_neutral} \\
&\phantom{\supset} +\frac{G_F}{\sqrt{2}} U_{ie}U^\dagger_{ej} \bar{e} \gamma_\mu (1- \gamma^5) e \, \bar{\nu}_i \gamma^\mu (1 - \gamma^5) \nu_j, \label{eq:L_nu_charged}
\end{align}
where the first and second lines contain the relevant neutral- and charged-current interactions, respectively, with $g^\text{V}_\psi$ and $g^\text{A}_\psi$ the vector and axial-vector couplings of the SM fermion $\psi$ to the $Z$ boson, and $U$ the unitary matrix that diagonalizes the neutrino mass matrix. 
(In this convention, 
$g^\text{V}_p = 1/2 - 2 s_w^2$, 
$g^\text{A}_p = 1/2$, 
$g^\text{V}_n = -1/2$,
$g^\text{A}_n = -1/2$,
$g^\text{V}_e = -1/2 + 2 s_w^2$,
$g^\text{A}_e = 1/2$, 
where $s_w$ is the sine of the weak mixing angle.)

Many of the oscillatory phenomena that can occur for linearly-coupled DM carry over to quadratic interactions through simple rescalings of the coupling and the oscillation frequency (now twice the mass), see e.g.~Refs.~\cite{Hees:2018fpg, Banerjee:2022sqg}. At higher masses, the leading effect becomes scattering in low-threshold targets rather than (double) absorption~\cite{Knapen:2017xzo}. When the field has no ambient density, there are still static forces from a virtual exchange of \emph{two} particles, but they decrease with distance as $1/r^4$ for two-scalar exchange rather than $1/r^2$ for single-scalar exchange, and is exponentially suppressed beyond half the Compton wavelength of the scalars~\cite{Ferrer:2000hm}. Famously, the feeble two-neutrino exchange forces scale as $G_F^2/r^6$~\cite{Feinberg:1968zz,Feinberg:1989ps,Sikivie:1983ip} at short distances.

In this paper, I calculate the \emph{density-dependent} forces between two SM particles or macroscopic bodies. If two SM particles are embedded in a sea of waves interacting with them via quadratic couplings, there exists a static force that scales linearly with the energy density, as the square of the quadratic couplings, and with a range of order the typical spatial (de Broglie) wavelength of the ambient waves, unless they have a high degree of coherence. The spatial profile of the resulting potentials is not generally spherically symmetric, and is determined by the couplings and the phase-space distribution of the mediating particles.  I call this phenomenon a \emph{wake force}, as it is analogous to the wake of a stationary boat (source particle) in a mild ocean swell (the ambient waves), which is felt by the nearby surrounding boats (the target particles). Curiously, wake forces generically violate Newton's third law---the vector sum of the wake force from source to target does not cancel that from target to source, since there is some momentum transferred (which depends on the source--target separation vector) to the ambient medium. 

In \Sec{sec:theory}, I present the general formalism for calculating (scalar, non-derivative) wake forces both classically (\Sec{sec:classical}) and quantum mechanically (\Sec{sec:quantum}), with analogous derivations for other scalar interactions and fermions relegated to \Apps{app:scalars} \&~\ref{app:fermions}. I calculate the range and spatial profile of wake forces in \Sec{sec:range}. In \Sec{sec:applications}, I discuss applications of wake forces to DM searches (\Sec{sec:dmsearches}) and neutrino detection~(\Sec{sec:neutrino}).  \Sec{sec:comparisons} contains parametric comparisons of wake forces and potentials to other effects that necessarily occur for quadratically-coupled fields: double-exchange forces (\Sec{sec:double_exchange}), elastic scattering~(\Sec{sec:elastic}), in-medium potentials and forces~(\Sec{sec:inmedium}), and screening~(\Sec{sec:screening}). 
A validation against nonperturbative numerical simulations is provided in \App{app:nonperturbative}.

Certain aspects of this work have appeared in prior literature in different guises, and were developed independently of the material presented here. I comment on similarities and differences with Refs.~\cite{Hees:2018fpg,Ferrer:1998ju,Ferrer:1999ad,Ferrer:2000hm,Berezhiani:2018oxf,Horowitz:1993kw,Ghosh:2022nzo,Arvanitaki:2022oby,Arvanitaki:2023fij,Blas:2022ovz} in \Sec{sec:discussion}, where I also discuss future directions and open questions.

Throughout this work, I use natural units with $\hbar = c = k_\mathrm{B} = 1$, the metric signature $(+,-,-,-)$, and conventions of $\int \ddbar^n k \equiv \int \dd^n k / (2\pi)^n$ and $\deltabar^{(n)}(\cdot) \equiv (2\pi)^n \delta^{(n)}(\cdot)$. Three-vectors are bolded ($\vect{k}$), four-vectors such as $k^\mu = (E_k,\vect{k}^2)$ and $x^\mu = (t,\vect{x})$ are not, and $k \cdot x = E_k t - \vect{k} \cdot \vect{x}$. I collect standard Fourier integrals and spinor identities in \App{app:formulae}.

\section{Theory} \label{sec:theory}

In this section, I develop the general framework for calculating the wake force between two particles, which can then be pair-wise integrated to obtain the force between two macroscopic bodies. The basic effect can be understood in simple classical terms, at the level of equations of motion and solutions thereof, presented in \Sec{sec:classical}. Alternatively, \Sec{sec:quantum} contains the exactly equivalent quantized treatment using tree-level scattering amplitudes, which may be illuminating vis-à-vis standard derivations from single-particle virtual exchanges, and is more powerful for spin-dependent interactions. 

The most minimal example is that of a real scalar field $\phi$ with a quadratic coupling to the number density current of a SM fermion $\psi = \lbrace p, n, e, \dots \rbrace$:
\begin{alignat}{2}
    \mathcal{L} = \half (\partial \phi)^2 - \frac{m^2}{2} \phi^2 - \frac{G m}{2} \phi^2 \bar{\psi} \psi. \label{eq:L_scalar}
\end{alignat}
The normalization of the interaction term proportional to both the $\phi$ mass $m$ and the coupling constant $G$ (in analogy to Fermi's constant) of inverse-mass-squared dimension is chosen for later notational convenience and to make contact with quadratically-coupled fermions. Ultraviolet (UV) completions of the EFT in \Eq{eq:L_scalar} are discussed in \Sec{sec:dmsearches}. The calculations in the rest of this section are repeated for other quadratic scalar interactions in \App{app:scalars} (e.g.~$|\Phi|^2 \bar{\psi} \psi$ for a complex scalar $\Phi$ and $\phi^2 \bar{\psi} i\gamma_5 \psi$) and for fermions in \App{app:fermions}. They follow the same steps as shown here.

\subsection{Classical Description} \label{sec:classical}

Consider a static ``source'' $\psi$ particle at the origin $\vect{x} = 0$, and a ``target'' $\psi$ particle at position $\vect{x}$ with $r \equiv |\vect{x}|$. The equation of motion for $\phi$ in the presence of this single source particle, with a number density of $\bar{\psi} \psi = \delta^{(3)}(\vect{x})$, is:
\begin{alignat}{2}
    (\Box + m^2)\phi = - G m \phi \deltathree(\vect{x}).
    \label{eq:EOM_scalar}
\end{alignat}
Assume that the $\phi$ medium is composed of a background wave $\phi_0(t,\vect{x})$ with amplitude $\varphi_0$ and wavenumber $\vect{k}_0$, and a scattered wave $\delta \phi(t,\vect{x})$:
\begin{alignat}{2}
    \phi &= \phi_0(t,\vect{x}) + \delta \phi(t,\vect{x}); \label{eq:phi_expansion}\\
    \phi_0(t,\vect{x}) &= \varphi_0 \cos(\omega t-\vect{k}_0\cdot \vect{x}) .
    \label{eq:phi_0}
\end{alignat}
The angular frequency is $\omega = \sqrt{m^2 + \vect{k}_0^2 }$ in vacuum; $\psi$-density-dependent corrections are discussed in \Sec{sec:screening}. \Sec{sec:quantum} will generalize the simple background plane wave of \Eq{eq:phi_0} to the more phenomenologically relevant case of a random superposition of different $\vect{k}$ modes.

\Eq{eq:EOM_scalar} can be solved perturbatively in $G$. Using the Green's function corresponding to the retarded propagator, one finds
\begin{alignat}{2}
    \delta \phi(t,\vect{x}) &= \frac{G m}{2} \varphi_0 \int  \ddbar^4p \, e^{-i p \cdot x} \frac{\deltabar(p_0-\omega) + \deltabar(p_0 + \omega)}{(p_0+i\varepsilon)^2-\vect{p}^2 - m^2 }\nonumber\\
    &= \frac{G m}{2} \varphi_0 e^{-i \omega t} \int  \ddbar^3p \, e^{i \vect{p}\cdot \vect{x}} \frac{1}{-\vect{p^2}+\vect{k}_0^2 + i\varepsilon}  + \mathrm{c.c.} \nonumber\\
    &=  -\frac{G m}{4\pi r} \varphi_0 \cos(\omega t-|\vect{k}_0|r) \label{eq:delta_phi}
\end{alignat}
as the leading-order solution in $G$ for the perturbed spherical wave $\delta \phi$. 

A ``target'' $\psi$ particle at $\vect{x}$ experiences the nonrelativistic potential:
\begin{alignat}{2}
    V &= \frac{G m}{2} \phi^2 \label{eq:V_1} \\
    &\simeq \frac{G m \left[\phi_0(t,\vect{x})\right]^2}{2} + G m \phi_0(t,\vect{x})\delta \phi(t,\vect{x}) + \mathcal{O}\left(G^3\right). \nonumber
\end{alignat}
The first term is the in-medium potential from the background wave only, and is independent of the source particle's position. The second term is the leading-order correction from the scattered wave. Its time-averaged expectation value is the wake potential:
\begin{align}
\langle V(t,\vect{x})\rangle_t = -\frac{G^2}{4\pi r}\frac{m^2 \varphi_0^2}{2} \cos(|\vect{k}_0|r-\vect{k}_0\cdot\vect{x}). \label{eq:V_2}
\end{align}
The wake force is the gradient of this potential, and is universally attractive for relative separations smaller than the de Broglie wavelength $r \ll 1/|\vect{k}_0|$. The parametric form of this wake potential,
\begin{alignat}{2}
    \hspace{-0.5em}
    V\sim - \frac{\text{[coupling]}^2 \times \text{[energy density]} \times \text{[form~factor]}}{4\pi \, \text{[radius]}}, \label{eq:V_heuristic}
\end{alignat}
is generic: it persists for other types of (spin-independent) quadratic interactions, including with complex scalars and fermions. The form factor is defined to go to unity in the small-radius and nonrelativistic limit. For phase-space distributions beyond that of a single $\vect{k}$ mode in \Eq{eq:V_2}, the form factor falls off significantly---polynomially or exponentially---at distances larger than the typical de Broglie wavelength. The form factor is also generally aspherical (at large distances), unless the phase-space distribution of $\phi$ particles is spherically symmetric itself. \Sec{sec:range} will cover the form factor in greater detail.

\begin{figure}[t]
    \includegraphics[width = 0.48\textwidth , trim = 0 0 0 0]{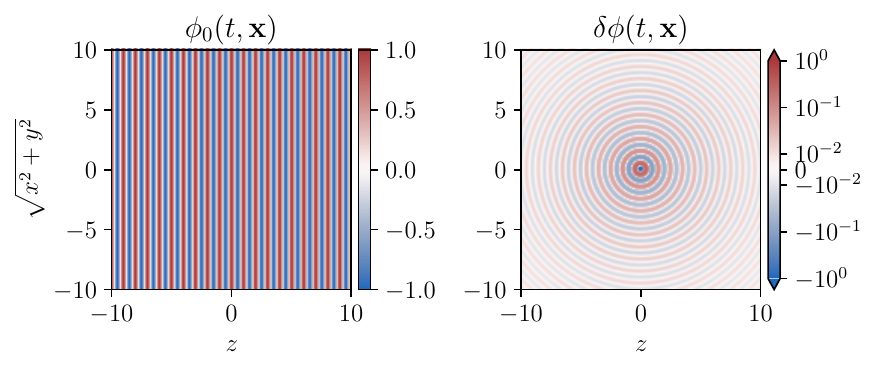}
    \includegraphics[width = 0.48\textwidth , trim = 0 0 0 0]{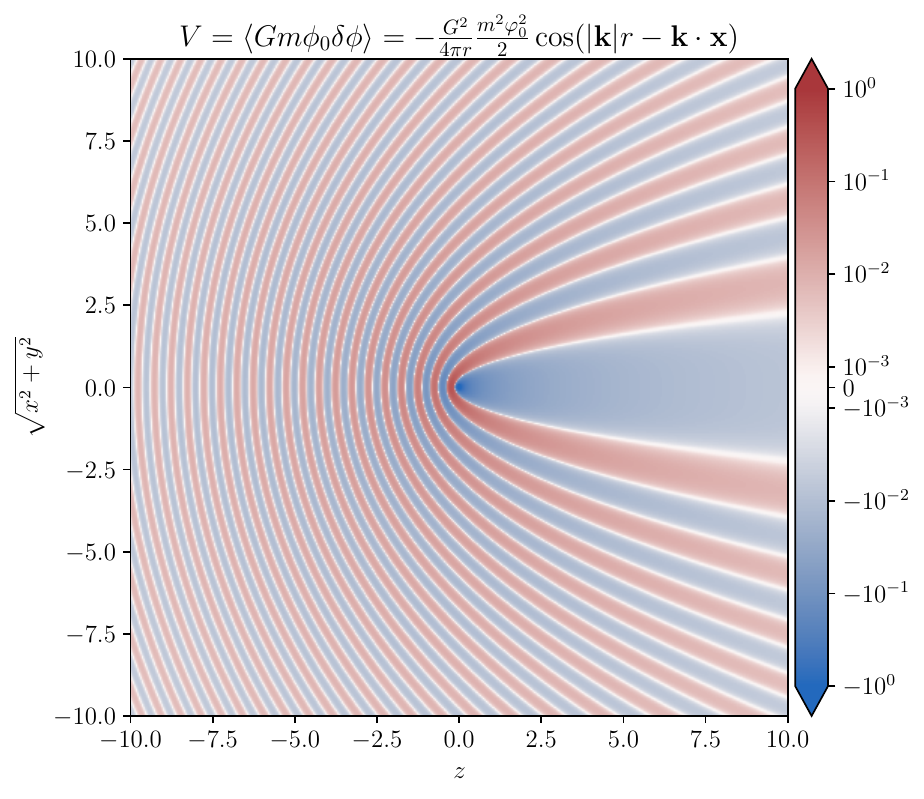}
    \caption{Time-averaged wake potential (bottom panel) of \Eq{eq:V_2} for a target test particle at position $\vect{x}$ produced by a monochromatic unidirectional background wave $\phi_0(t=0,\vect{x})$ of \Eq{eq:phi_0} moving in the $+z$ direction (top left panel) and the scattered spherical wave $\delta \phi(t=0,\vect{x})$ of \Eq{eq:delta_phi} (top right panel) from a perturbing source particle at the origin. For illustrative purposes, units are such that $G = m = \varphi_0 = 1$ and $\vect{k} = (0,0,2\pi)$.}\label{fig:wake}
\end{figure}

\Fig{fig:wake} depicts the wake potential (bottom panel) for a monochromatic unidirectional wave moving in the $z$-direction $\hat{\vect{k}} = (0,0,2\pi)$, as the multiplicative cross-term (the second term on the RHS of \Eq{eq:V_1}) of the background wave (top left panel) and the perturbed spherical wave (top right panel). For the \emph{monochromatic unidirectional} background wave of \Eq{eq:phi_0}, the wake force effectively has an infinite range in the forward direction $\hat{\vect{x}} \simeq \hat{\vect{k}}$, and is spatially oscillatory with wavenumber magnitude $\mathcal{O}(|\vect{k}|)$ at large angles ($2 |\vect{k}|$ in the backward direction). The generalization to a random superposition of background waves of varying wavenumber magnitudes and directions is crucial for phenomenological purposes, and is more easily derived in the quantized treatment below.

\subsection{Quantum Description} \label{sec:quantum} 

The classical calculation from the previous section can be repeated from the perspective of tree-level scattering amplitudes in quantum field theory, where the generalization to an ensemble of many $\vect{k}$ modes and other types of interactions (\Apps{app:scalars} \&~\ref{app:fermions}) is more transparent. The real scalar field of \Eq{eq:L_scalar} is quantized in the usual way:
\begin{alignat}{2}
    \phi = \int \frac{\ddbar^3k}{\sqrt{2E_k}} \left[ a_{\vect{k}} e^{-i k\cdot x} + a_{\vect{k}}^\dagger e^{+ik\cdot x} \right], \label{eq:phi_quant}
\end{alignat}
with $E_k^2 = m^2 + \vect{k}^2$. 
The environment, i.e.~the ensemble of background waves, is taken to be a mixed state of $\vect{k}$ modes, with a momentum distribution $f(\vect{k})$ normalized as $\int \ddbar^3k \, f(\vect{k}) = 1$. In this background, the expectation value of the product of creation and annihilation operators is:
\begin{align}
    \left\langle a_{\vect{k}'}^\dagger a_{\vect{k}} \right\rangle =  n f(\vect{k}) \deltabarthree(\vect{k}'-\vect{k}), \label{eq:exp_mixed}
\end{align}
with $n$ the number density of $\phi$ particles. Using the free-particle Hamiltonian density $\mathcal{H} = [\dot{\phi}^2 + (\vect{\nabla}\phi)^2 + m^2 \phi^2 ]/2$, the expectation value of the energy density is $\rho = \langle \mathcal{H} \rangle = n \int \ddbar^3k \, E_k f(\vect{k})$. 

\begin{figure}[t]
\begin{tikzpicture}
    \begin{feynman}
      \vertex (i1) {\(\psi_1\)};
      \vertex [above=of i1] (v1);
      \vertex [above=of v1] (f1) {\(\psi_1\)};
      \vertex [right=of i1] (i2) {\(\psi_2\)};
      \vertex [above=of i2] (v2);
      \vertex [above=of v2] (f2) {\(\psi_2\)};
      \vertex [left=1cm of i1]  (is) {\(\phi\)};
      \vertex [right=1cm of f2] (fs) {\(\phi\)};
      \diagram* {
      (i1) -- [fermion, very thick] (v1) -- [fermion, very thick] (f1),
      (i2) -- [fermion, very thick] (v2) -- [fermion, very thick] (f2),
      (is) -- [scalar] (v1) -- [scalar, edge label=\(\phi\), momentum'=\(q+k\)] (v2) -- [scalar] (fs)
      };
    \end{feynman}
  \end{tikzpicture} 
  \quad
  \begin{tikzpicture}
    \begin{feynman}
      \vertex (i1) {\(\psi_1\)};
      \vertex [above=of i1] (v1);
      \vertex [above=of v1] (f1) {\(\psi_1\)};
      \vertex [right=of i1] (i2) {\(\psi_2\)};
      \vertex [above=of i2] (v2);
      \vertex [above=of v2] (f2) {\(\psi_2\)};
      \vertex [right=1cm of i2]  (is) {\(\phi\)};
      \vertex [left=1cm of f1] (fs) {\(\phi\)};
      \diagram* {
      (i1) -- [fermion, very thick] (v1) -- [fermion, very thick] (f1),
      (i2) -- [fermion, very thick] (v2) -- [fermion, very thick] (f2),
      (is) -- [scalar] (v2) -- [scalar, edge label'=\(\phi\), reversed momentum=\(q-k\)] (v1) -- [scalar] (fs)
      };
    \end{feynman}
  \end{tikzpicture} 
  \caption{Diagrams contributing to the wake force between distinguishable, nonrelativistic fermions $\psi_1$ and $\psi_2$ in a medium of real scalars $\phi$.}\label{fig:feyn_scalar}
\end{figure}
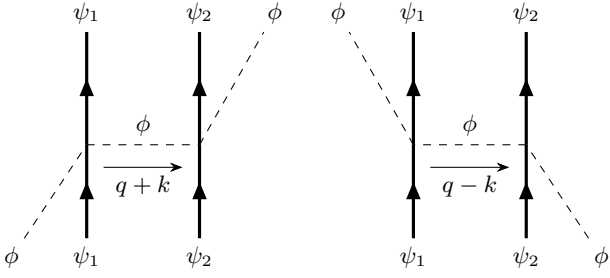

The Feynman diagrams contributing to the wake force are depicted in \Fig{fig:feyn_scalar}. Taking the nonrelativistic limit of distinguishable SM fermions $\psi_1$ and $\psi_2$ with masses $m_\psi$, the matrix element $\mathcal{M}$ for this scattering amplitude is
\begin{align}
    &\frac{i\mathcal{M}}{(2m_\psi)^2} = \label{eq:matrix_scalar}\\
    &- i G^2 m^2 \int \ddbar^3k\, \frac{n  f(\vect{k})}{2E_k} \left[\frac{1}{(q+k)^2-m^2}+ \frac{1}{(q-k)^2 - m^2} \right] \nonumber
\end{align}
in the background of \Eq{eq:exp_mixed}. The four-momentum exchanged between $\psi_1$ and $\psi_2$ is denoted by $q$ with direction as indicated in \Fig{fig:feyn_scalar}. Use of the standard relation $\widetilde{V}(\vect{q}) = - \mathcal{M}/(2m_\psi)^2$ between the matrix element and the 3D Fourier transform of the potential $\widetilde{V}(\vect{q}) = \int \dd^3x \, e^{-i\vect{q}\cdot\vect{x}} V(\vect{x})$ yields the result:
\begin{alignat}{2}
    \widetilde{V}(\vect{q}) &= - G^2 m^2\int \ddbar^3k\, \frac{n  f(\vect{k})}{2E_k} \label{eq:V_3}\\
&\phantom{=} \hspace{2em}\times \left[ \frac{1}{\vect{q}^2+2\vect{q}\cdot \vect{k}-i\varepsilon} + \frac{1}{\vect{q}^2-2\vect{q}\cdot \vect{k}+i\varepsilon} \right],\nonumber
\end{alignat}
where, once again, the $i \varepsilon$ prescription corresponds to the use of the retarded propagator to retain causality. Fourier transforming back to position space gives the wake potential:
\begin{align}
    V(\vect{x}) & = \int \ddbar^3q\, e^{i\vect{q}\cdot\vect{x}} \widetilde{V}(\vect{q}) \nonumber \\
    & = -G^2 m^2 \int \ddbar^3k\,  \frac{n  f(\vect{k})}{2E_k} \int\ddbar^3q\, \frac{e^{i(\vect{q}-\vect{k})\cdot\vect{x}}}{\vect{q}^2-\vect{k}^2-i\varepsilon} + \cc \nonumber \\
    & = -\frac{G^2 m^2}{4\pi r} \int \ddbar^3k\,  \frac{n  f(\vect{k})}{E_k} \cos(|\vect{k}|r - \vect{k}\cdot\vect{x}). \label{eq:V_4}
\end{align}
In the second line, the integration variable is shifted as $\vect{q}+\vect{k} \to \vect{q}$. The result in the third line is equivalent to the classical result of \Eq{eq:V_2} for $f(\vect{k}) = \deltabarthree(\vect{k}-\vect{k}_0)$ and with the identification $n/E_{k_0} = \varphi_0^2/2$. 

The long range of the wake force (of order the de Broglie wavelength) is explained by the nearly on-shell internal line: the square of the four-momentum $k$ carried by the on-shell external state cancels the pole in the internal propagator in \Fig{fig:feyn_scalar} and \Eq{eq:matrix_scalar}. In contrast, the short range of Yukawa-type forces (of order the Compton wavelength) from single-particle exchange and of double-particle-exchange forces can be attributed to their off-shell internal lines.

The parametric form promised in \Eq{eq:V_heuristic},
\begin{align}
    \boxed{V(\vect{x}) = -\frac{G^2 m n }{4\pi r}  \mathcal{F}(\vect{x}), \label{eq:V_wake_scalar}}
\end{align}
is recovered in the nonrelativistic limit, where $\lim_{\vect{x}\to 0}\mathcal{F}(\vect{x}) \simeq 1$ and $\rho \simeq m n $. This is the relevant limit for the DM searches in \Sec{sec:dmsearches}, and will be used to calculate the range of the wake force in the next section. 

\subsection{Range} \label{sec:range}

The range and profile of the wake force is governed by the momentum distribution $f(\vect{k})$ of the background particles, which determines the form factor in \Eq{eq:V_wake_scalar}:
\begin{alignat}{2}
    \boxed{\mathcal{F}(\vect{x}) \equiv \int \ddbar^3k\,  \frac{f(\vect{k})}{E_k/m} \cos(|\vect{k}|r - \vect{k}\cdot\vect{x}) \label{eq:form_general}.}
\end{alignat}

\begin{figure}[t]
    \includegraphics[width = 0.48\textwidth , trim = 0 0 0 0]{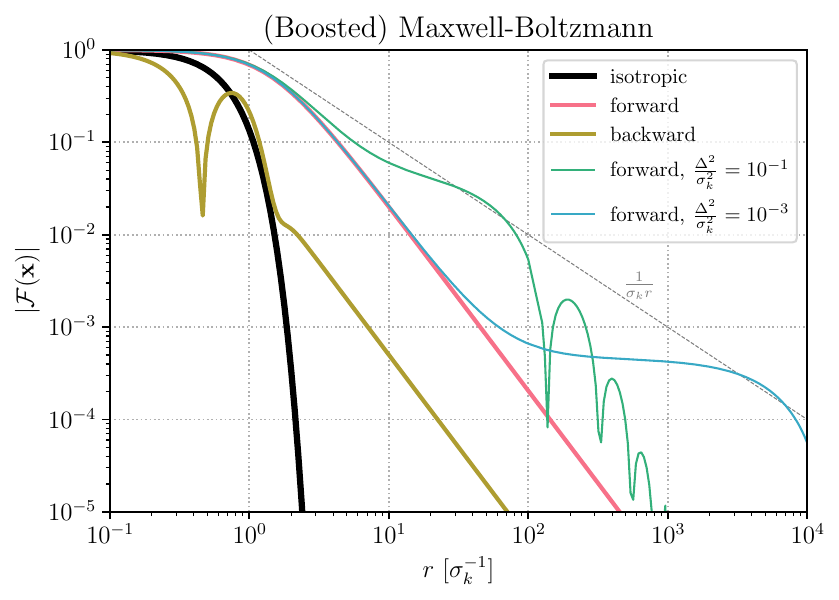}
    \includegraphics[width = 0.48\textwidth , trim = 0 0 0 0]{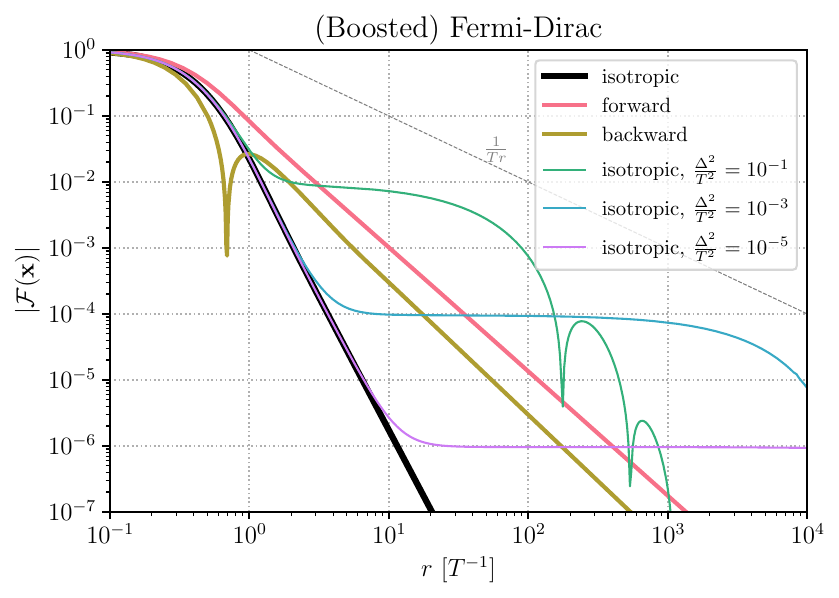}
    \caption{
    Distance ($r$) dependence of the form factor magnitude $|\mathcal{F}(\vect{x})|$ for MB (top panel) and FD (bottom panel) momentum distributions. The black curves are the isotropic form factors of \Eqs{eq:form_MB} \& \ref{eq:form_FD} (no boost). The red (gold) curves are for BMB and BFD distributions of \Eqs{eq:mom_BMB} \&~\ref{eq:mom_BFD} in the forward (backward) direction. The other colored curves (green, blue, purple) are for small mass-squared splittings $\Delta^2 > 0$, in the forward direction for the BMB distribution, and for the isotropic FD distribution. The boost/anisotropy alone lifts the form factor suppression to $F\propto 1/r^2$ in all cases, while for small mass splittings, $F \sim \Delta^2 / |\vect{k}_0|^2$ for $r \lesssim \lambda_\mathrm{osc} \sim |\vect{k}_0| / |\Delta^2|$ can be achieved, where $|\vect{k}_0|$ is the typical momentum in the distribution.
    }\label{fig:form_radial}
\end{figure}

\paragraph*{Isotropic distributions---}
For an isotropic Maxwell-Boltzmann (MB) distribution, the form factor takes on a simple analytic form (for $E_k/m \simeq 1$):
\begin{align}
f_\mathrm{MB}(\vect{k}) &= \frac{(2\pi)^{3/2}}{\sigma_{k}^3} e^{-\vect{k}^2/2\sigma_k^2}, \label{eq:mom_MB} \\
\mathcal{F}_\mathrm{MB}(\vect{x}) &\simeq  e^{-2\sigma_k^2 r^2}; \label{eq:form_MB}
\end{align}
falling off as a (spherical) Gaussian beyond the characteristic de Broglie wavelength $\sigma_k^{-1}$. It is depicted as the thick black curve in the top panel of \Fig{fig:form_radial}. For an isotropic Fermi-Dirac (FD) distribution with temperature $T$:
\begin{align}
f_\mathrm{FD}(\vect{k}) &= \frac{(2\pi)^2}{3\zeta(3)}\frac{1}{e^{|\vect{k}|/T}+1} \frac{1}{T^3}, \label{eq:mom_FD} \\
\mathcal{F}_\mathrm{FD}(\vect{x}) &= -i\frac{\psi^{(1)}(1/2-i T r) - \psi^{(1)}(1-i T r)}{24 \zeta(3) T r} + \cc \nonumber \\
&\simeq \frac{1}{48\zeta(3) (T r)^4} + \mathcal{O}(Tr)^{-6} \quad (r \gg 1/T);
& \label{eq:form_FD}
\end{align}
where $\psi^{(1)}$ is the first derivative of the digamma function.
In this case, the form factor is quartically suppressed for $r \gg 1/T$, shown as the thick black curve in the bottom panel of \Fig{fig:form_radial}.

For a general isotropic momentum distribution $f(\vect{k}) = f(|\vect{k}|)$, the form factor integral reduces to:
\begin{alignat}{2}
    \mathcal{F}(r) = \frac{1}{(2\pi)^2 r} \int_0^\infty \dd k \, k \sin( k r) f(k) \label{eq:form_isotropic},
\end{alignat}
proving at least $1/r$ suppression at large $r$, and even more suppression for smooth phase-space distributions due to the oscillatory nature of the integrand. For a (physically less motivated) top-hat momentum distribution, $f(\vect{k}) = 6\pi^2 k_0^{-3} \Theta(k_0 - |\vect{k}|)$, the form factor decreases in an oscillatory fashion with an inverse quadratic envelope, according to the parametric form $\mathcal{F} \propto \cos(2k_0r)/(k_0 r)^2$ for $r \gg 1/k_0$.  The top-hat distribution shows that despite ``hard edges'' in $f(\vect{k})$, there is still a significant suppression of the form factor beyond the typical wavelength $k_0^{-1}$. (These edges may come from incomplete phase-space equilibration at velocities below the solar system's escape velocity, or from the screening effects discussed in \Sec{sec:screening}.)

\paragraph*{Anisotropic distributions---}
The momentum distributions for many background fields of phenomenological interest, such as the DM and C$\nu$B particles, are strongly anisotropic in the laboratory frame due to our motion relative to that of the MW DM halo or the cosmic rest frame. 

\begin{figure}[t]
    \includegraphics[width = 0.48\textwidth , trim = 0 0 0 0]{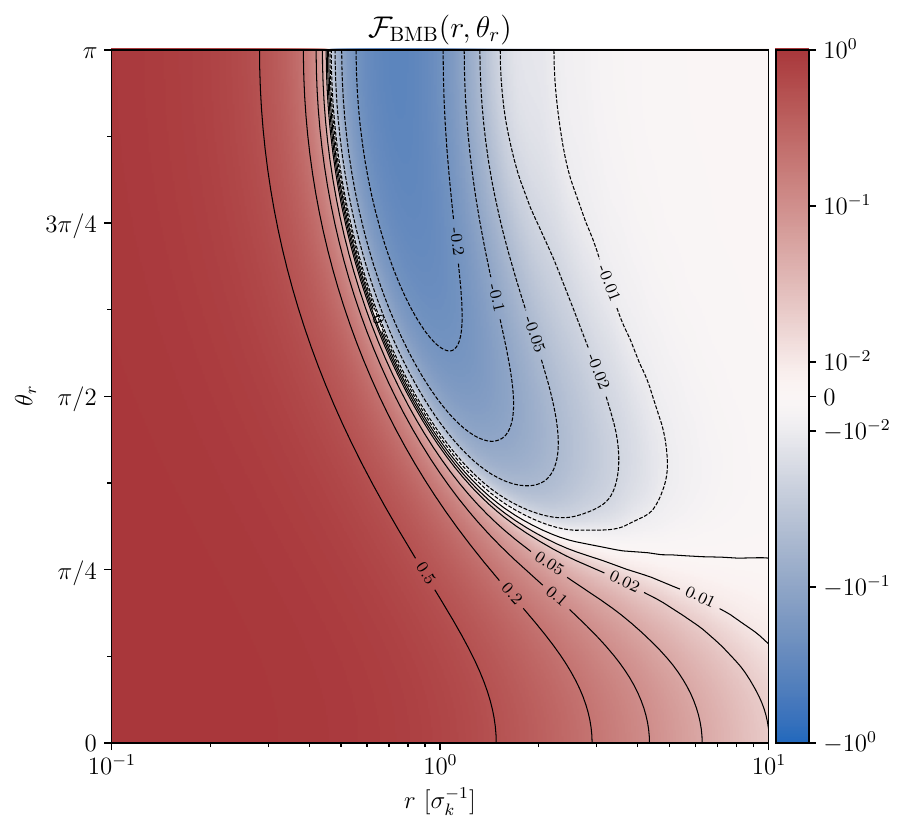}
    \includegraphics[width = 0.48\textwidth , trim = 0 0 0 0]{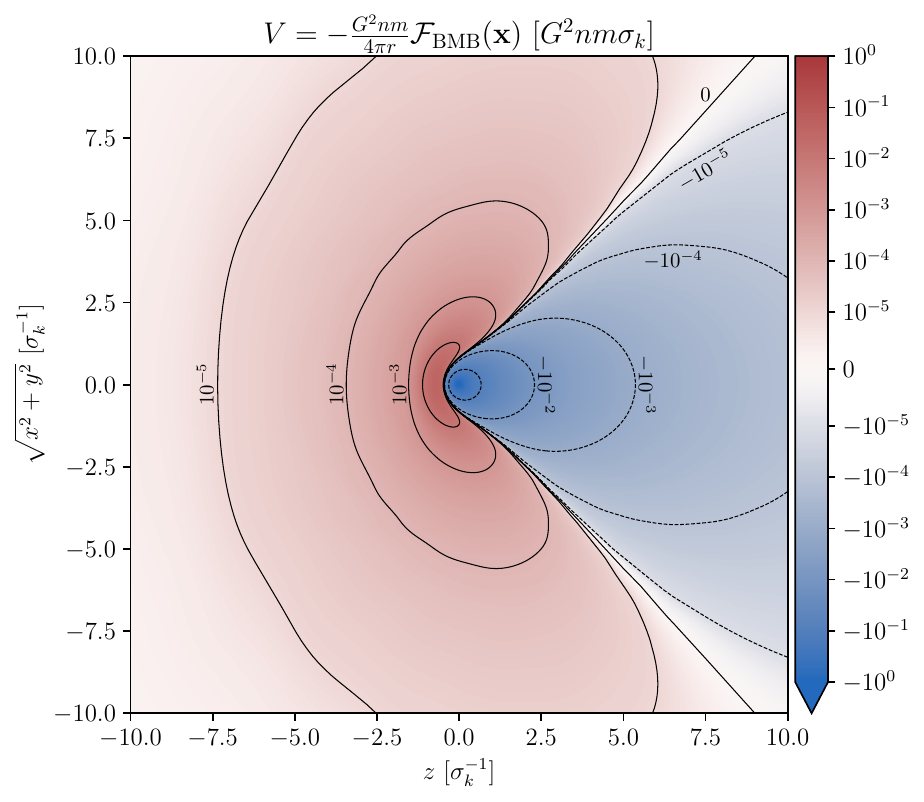}
    \caption{
        \textit{Top panel:} Form factor $\mathcal{F}_\mathrm{BMB}(\vect{x})$ for a boosted Maxwell-Boltzmann distribution at distance $r$ and angle $\theta_r \equiv \arccos(\hat{\vect{x}} \cdot \hat{\vect{z}})$ relative to the direction of the DM wind. \textit{Bottom panel:} Wake potential in units of $G^2 n m \sigma_k / 2$ to allow for direct comparison with the bottom panel of \Fig{fig:wake}. At large radii, $V_\mathrm{BMB} \propto 1/r^3$ in all directions. 
    }\label{fig:wake_BMB}
\end{figure}

\begin{figure}[t]
    \includegraphics[width = 0.48\textwidth , trim = 0 0 0 0]{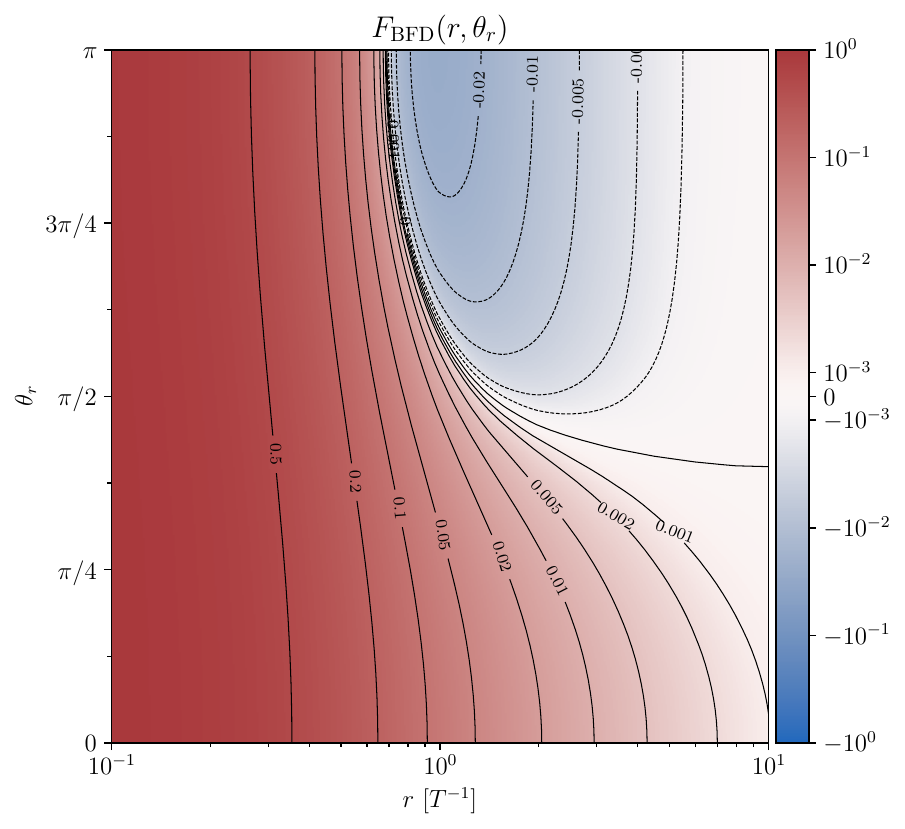}
    \includegraphics[width = 0.48\textwidth , trim = 0 0 0 0]{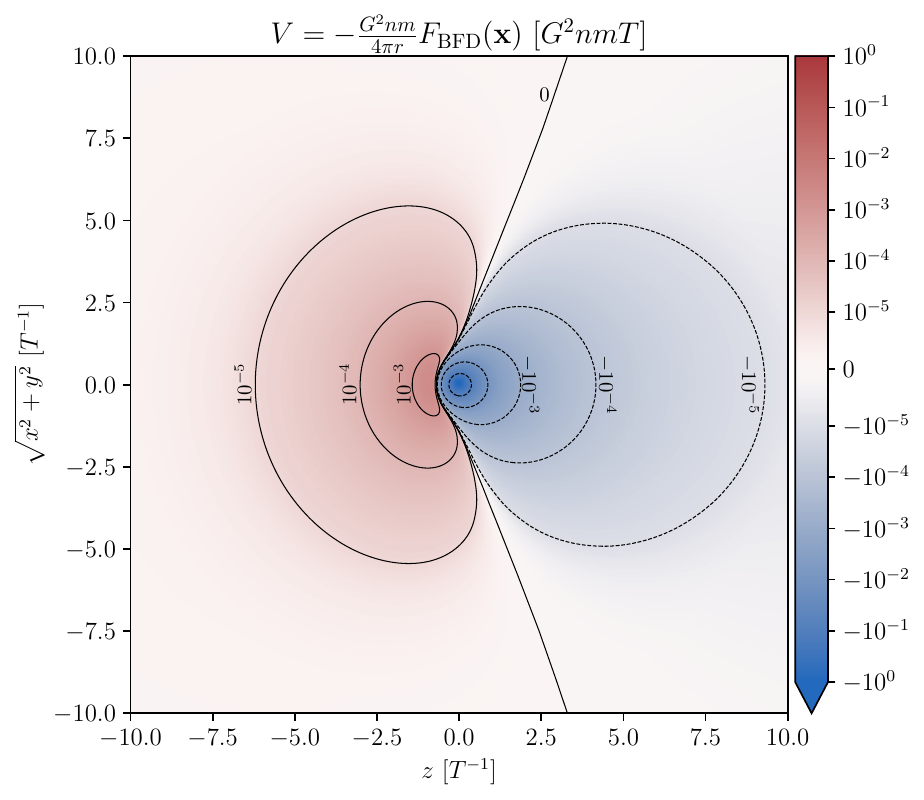}
    \caption{
        Same as in \Fig{fig:wake_BMB}, but for a boosted Fermi-Dirac distribution for $m = 60\,\mathrm{meV}$, $T_\nu = 1.95\,\mathrm{K}$, and $v_\mathrm{CMB} = 370\,\mathrm{km/s}$ pointing in the $+z$-direction. At large radii, $V_\mathrm{BFD} \propto 1/r^3$ in all directions.
    }\label{fig:wake_BFD}
\end{figure}

The DM halo in Earth's frame can be modeled to first approximation by a truncated, boosted Maxwell-Boltzmann (BMB) momentum distribution:
\begin{align}
f_\mathrm{BMB}(\vect{k}) &= \mathcal{N}_\mathrm{esc}^{-1} f_\mathrm{MB}(\vect{k}-m v_\mathrm{circ} \hat{\vect{z}}) \Theta(m v_\mathrm{esc} - |\vect{k}|), \label{eq:mom_BMB}
\end{align}
with coordinates chosen such that the DM wind points towards the $+z$-direction (circular velocity $\vect{v}_\mathrm{circ} = -v_\mathrm{circ} \hat{\vect{z}}$). The distribution is truncated above the escape velocity $v_\mathrm{esc}$, which affects the normalization via $\mathcal{N}_\mathrm{esc} = \mathrm{erf}(\bar{z}) - 2\bar{z} e^{-\bar{z}^2}/\sqrt{\pi}$, where $\bar{z} \equiv m v_\mathrm{esc}/\sqrt{2}\sigma_k$. 
The corresponding form factor $\mathcal{F}_\mathrm{BMB}$ cannot be calculated analytically. However, its numerical evaluation is shown in the top panels of \Fig{fig:wake_BMB} and \Fig{fig:form_radial} for fiducial parameters for the velocity distribution at the Sun's location in the Milky Way (MW): $\sigma_v \equiv \sigma_k/m \approx 165\,\mathrm{km/s} \approx \sqrt{2} v_\mathrm{circ}$ and $v_\mathrm{esc} \approx 550.7\,\mathrm{km/s}$~\cite{Catena:2011kv}. At distances larger than those shown in \Fig{fig:wake_BMB}, the form factor falls off \emph{quadratically} in all directions. The distance dependence in the forward (backward) direction is indicated by the red (gold) curve in \Fig{fig:form_radial}. The exponential suppression for the isotropic MB distribution is thus lifted, but it still constitutes a dramatic suppression relative to the forward limit from \Eq{eq:V_2}---compare the bottom panels of \Figs{fig:wake} \&~\ref{fig:wake_BMB}. 

In a standard cosmology, the C$\nu$B is expected to be at rest relative to the cosmic microwave background (CMB) at a temperature $T_\nu \simeq (4/11)^{1/3} T_\gamma \approx 1.95\,\mathrm{K}$, where $T_\gamma \approx 2.725\,\mathrm{K}$ is the CMB temperature~\cite{Fixsen:2009ug}. The CMB dipole anisotropy thus implies that the C$\nu$B is moving with respect to the Earth at a velocity of $v_\mathrm{CMB} \approx 370\,\mathrm{km/s}$ in the direction of the CMB dipole~\cite{Planck:2018nkj}. The corresponding momentum distribution is a boosted Fermi-Dirac (BFD) distribution:
\begin{align}
f_\mathrm{BFD}(\vect{k}) &= f_\mathrm{FD}(\vect{k}-m v_\mathrm{CMB} \hat{\vect{z}}). \label{eq:mom_BFD}
\end{align}
\Fig{fig:wake_BFD} displays the form factor $\mathcal{F}_\mathrm{BFD}$ for the C$\nu$B in Earth's frame for $m = m_3 = 60\,\mathrm{meV}$, a fiducial value for the largest of the three neutrino masses, in agreement with current constraints on the sum of neutrino masses~\cite{Simpson:2017qvj} and not far from the lowest possible value allowed by neutrino oscillation experiments~\cite[Ch.~14]{ParticleDataGroup:2022pth}. The strength of the anisotropy is quantified by $m v_\mathrm{CMB} / T_\nu$, which equals about $0.44$ for the assumed mass $m_3$, and is significantly smaller for the lighter mass eigenstates. The form factor is again suppressed relative to the forward limit from \Eq{eq:V_2}. Like for the BMB distribution, the boost mitigates the distance suppression of the form factor to $\mathcal{F}_\mathrm{BFD} \propto 1/r^2$ at large radii, which is less severe than the $1/r^4$ scaling of the unboosted $\mathcal{F}_\mathrm{FD}$---cfr.~the bottom panel of \Fig{fig:form_radial}.

The reason for the $1/r^3$ scaling of the wake potential is analogous to the $1/r^3$ effective potential of a charged particle in a Vlasov plasma with either a relative velocity or anisotropy~\cite{Montgomery:1968}. In that case too, the boost and/or anisotropy lift the exponential Debye screening of the potential to a polynomial (cubic) one.

The anisotropic momentum distributions $f_\mathrm{BMB}$ and $f_\mathrm{BFD}$ are but rudimentary approximations to the actual phase-space distributions of DM and the C$\nu$B, respectively. Effects that would certainly alter the DM form factor include: MW substructure such as streams and subhalos, the annual modulation from Earth's orbit around the Sun, environmental screening (\Sec{sec:screening}), and corrections to the phase-space distribution at low velocities ($|\vect{k}|/m \lesssim 10^{-4}$) due to the gravitational potential of the Sun. Similarly, the momentum distribution of the heaviest neutrino mass eigenstate likely deviates from a BFD distribution due to its gravitational clustering within the Virgo/Laniakea Supercluster, and, to a lesser extent, the MW halo and solar system. A detailed study of these corrections is left to future work.

\paragraph*{Multiple fields---} The form factor can parametrically change for \emph{multiple} wake fields, possibly extending the range of the wake force through a phenomenon akin to neutrino oscillations. Suppose the Lagrangian of \Eq{eq:L_scalar} is modified to an interaction with two real scalar fields $\phi_1$ and $\phi_2$, with masses $m_1$ and $m_2$:
\begin{align}
\mathcal{L} = - G m_1 \phi_1 \phi_2 \bar{\psi} \psi.
\end{align}
For concreteness, assume there is only an ambient $\phi_1$ field with number density $n$ and momentum distribution $f(\vect{k})$, and that $\phi_2$ is in its vacuum state. If $m_1 = m_2 = m$, one would get exactly the same results as for a single real scalar field of mass $m$, from the same diagrams as in \Fig{fig:feyn_scalar}, except with $\phi_2$ on the internal line, and $\phi_1$ on the external legs.

Qualitatively new dynamics occur for a mass splitting:
\begin{align}
\Delta^2 = m_2^2 - m_1^2. \label{eq:Delta}
\end{align}
If $|\Delta|$ is greater than the typical momentum $|\vect{k}_0|$ in $f(\vect{k})$, then the wake force has a shorter range, of order $1/|\Delta|$ instead of $1/|\vect{k}_0|$, because the propagator is further off shell.
More interesting is the opposite limit: $0 < |\Delta^2| \ll \vect{k}_0^2$. First, the appropriate generalization of the form factor is:
\begin{align}
\mathcal{F}(\vect{x},  \Delta^2) \equiv \text{Re}\int \ddbar^3 k\, \frac{f(\vect{k})}{\sqrt{1+\frac{\vect{k}^2}{m_1^2}}} e^{i(\sqrt{\vect{k}^2-\Delta^2}r-\vect{k}\cdot\vect{x})}. \label{eq:form_osc}
\end{align}
At distances $r \lesssim 1/|\vect{k}_0|$, the form factor is almost the same as if $\Delta^2$ were to vanish exactly.

To illustrate the oscillation phenomenon most simply, consider an isotropic momentum distribution $f(\vect{k}) = f(|\vect{k}|)$:
\begin{align}
\mathcal{F}(r,  \Delta^2) &= \frac{1}{2\pi^2 r} \text{Re} \int_0^\infty \dd k\, \frac{k f(k) }{\sqrt{1+\frac{k^2}{m_1^2}}}e^{i\sqrt{k^2-\Delta^2}r}\sin(k r)\nonumber\\
&\simeq \frac{1}{4\pi^2} \frac{1}{r} \int_0^\infty \dd k\, \frac{k f(k) }{\sqrt{1+\frac{k^2}{m_1^2}}} \sin\left(\frac{\Delta^2r}{2k} \right), \label{eq:form_osc_1}
\end{align}
with the approximation in the second line valid for $|\vect{k}_0| r \gg 1$ and $0<|\Delta| / |\vect{k}_0| \ll 1$, with $|\vect{k}_0|$ again the typical momentum in $f(|\vect{k}|)$. A new length scale appears in the form factor:
\begin{align}
\lambda_\text{osc} \sim \frac{|\vect{k}_0|}{|\Delta^2|}, \label{eq:osc}
\end{align}
which is the inverse of the spatial ``beat'' wavenumber of the two oscillatory factors in the first line of \Eq{eq:form_osc_1}, and is reminiscent of  neutrino oscillations. As in that case, the oscillation length scale arises because of the mismatch of interaction and mass eigenstates, and is parametrically equal to the characteristic wavenumber divided by the mass-squared splitting.
Thus, instead of being severely suppressed for $r\gg 1/|\vect{k}_0|$, the form factor generally takes on a size
\begin{align}
\mathcal{F}(r,\Delta^2) \sim \frac{\Delta^2}{|\vect{k}_0|^2} \equiv \mathcal{F}_\text{osc} \quad (1/|\vect{k}_0| \ll r \lesssim \lambda_\text{osc}),
\end{align}
largely independent of the precise shape of the momentum distribution. In a sense, the form factor is just linearly suppressed at $r \sim \lambda_\mathrm{osc}$, since $\mathcal{F}(\lambda_\mathrm{osc},  \Delta^2) \sim \mathrm{sgn}(\Delta^2)/(|\vect{k}_0|\lambda_\mathrm{osc} )$, as visually indicated by the gray dashed lines in \Fig{fig:form_radial}. Note that the sign of the form factor---and thus the attractive or repulsive nature of the wake force---is commensurate with the sign of $\Delta^2$ over this distance range. The behavior for various ratios of $|\Delta| / \sigma_k$ and $|\Delta|/T$ for $\Delta^2 > 0$ is plotted in \Fig{fig:form_radial}; other than a sign change at large distances, the negative-$\Delta^2$ case is similar.

\section{Applications} \label{sec:applications}

\subsection{Dark Matter Searches} \label{sec:dmsearches}
In this section, I explore the application of wake forces to searches for DM. As a strawman example for a first case study, I focus on the quadratic scalar coupling to nucleons $N = (p,n)$:
\begin{alignat}{2}
    \mathcal{L} \supset -\frac{G_{s,N}}{2} m \phi^2 \bar{N} N, \label{eq:G_s_N}
\end{alignat}
though the generalization to the analogous electron coupling $-(G_{s,e}/{2}) m \phi^2 \bar{e} e$ is straightforward. (In particular, given that the number density of nucleons is roughly twice that of electrons in standard matter, one can rescale the resulting wake force from the nucleon coupling to the total coupling $G_{s,\mathrm{tot}} \cong G_{s,N} + G_{s,e}/2$ to first approximation.)

The low-energy EFT may also include a parity-violating interaction with the pseudoscalar current $-(G_{p,N}/2) m \phi^2 \bar{N} i\gamma^5 N$, analogous (derivative) couplings to the pseudovector current, or similar parity-violating interactions with electron currents. 
These will lead to spin-dependent dipole-dipole wake forces proportional to $G_p^2$, or in combination with the quadratic scalar-current couplings of \Eq{eq:G_s_N}, to spin-dependent monopole-dipole wake forces proportional to $G_s G_p$, as calculated in \App{app:scalars}. 
Their employment in the search for DM is fruit for further studies.

As shown in \App{app:fermions}, quadratic fermionic couplings of the form $-(G_{s,N}/2) \bar{\chi} \chi \bar{N} N$ lead to precisely the same wake force as \Eq{eq:G_s_N} in the nonrelativistic limit, so in principle wake forces can be used to search for fermionic DM particles $\chi$ as well. However, the relatively short range of the wake force (while at least $10^3$ times longer than that of the double-exchange force of \Sec{sec:double_exchange} in vacuum) precludes its practical use in searches for realistic fermionic DM candidates, due to the lower bound on the mass: $m_\chi \gtrsim \mathrm{keV}$~\cite{Tremaine:1979we}. This confines the (unsuppressed) wake force range to the sub-micron regime, where current experiments are less sensitive.

\begin{figure*}[t]
\centering
\includegraphics[width=0.98\textwidth]{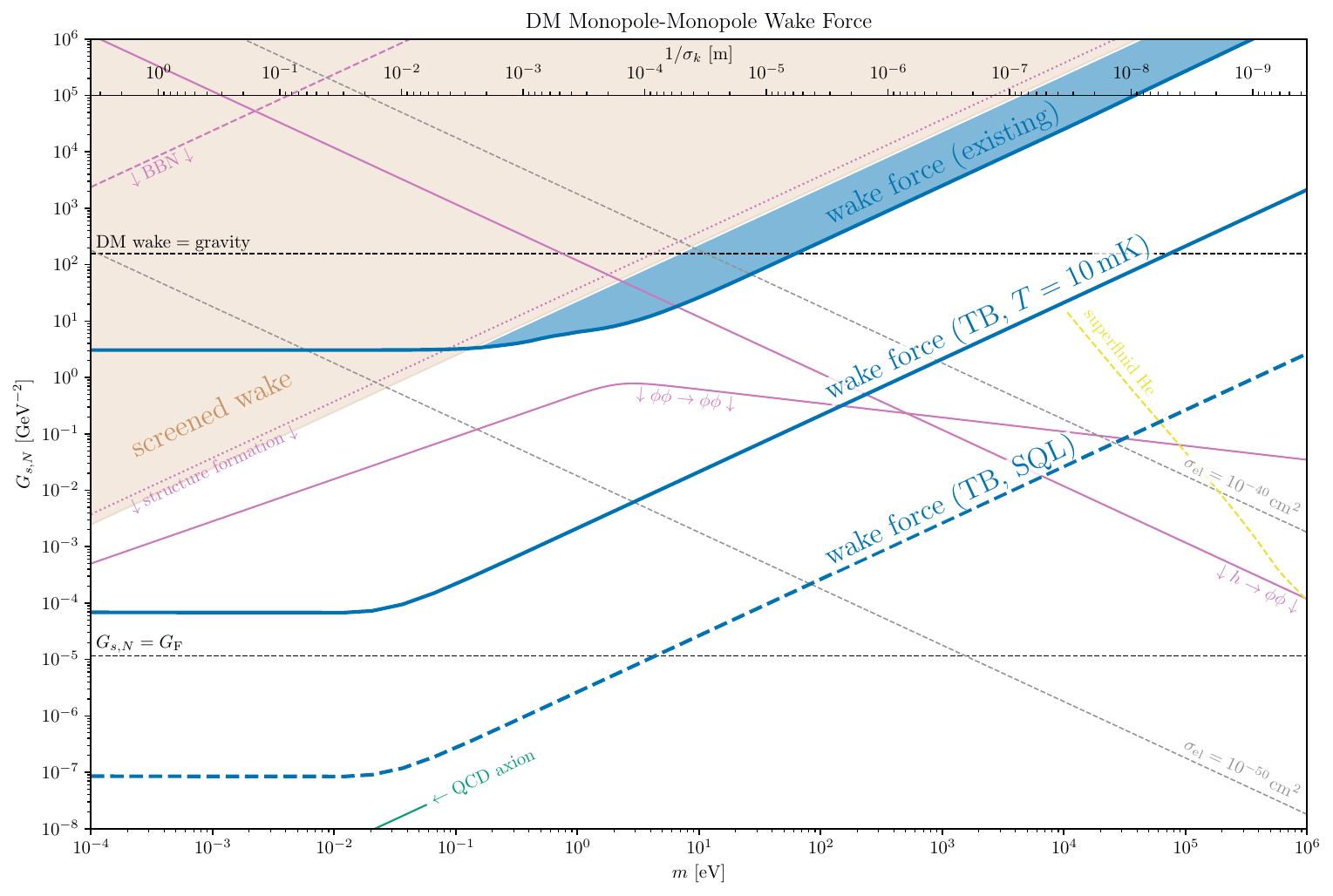}
\caption{Parameter space of quadratic scalar coupling $G_{s,N}$ to nucleons as a function of $\phi$ mass $m$, in the interaction $\mathcal{L} \supset -(G_{s,N} m /2) \phi^2 \bar{N} N$. In the blue region, the wake force is within reach of existing monopole-monopole force experiments if $\phi$ makes up all the DM. The other blue lines indicate wake forces detectable by a torsion balance (TB) with $1\,\mathrm{cm}$-radius tungsten spheres as source and target masses, angular frequency $\omega = 2\pi \,\mathrm{mHz}$, quality factor $Q = 10^8$, and noise temperatures $T = 10\,\mathrm{mK}$ (solid) and $T = \omega/2$ (dashed, SQL). In the light brown region, the wake force is screened by matter at standard densities. The characteristic range $1/\sigma_k$ is indicated by the top axis. Below the upper black dashed line, the DM wake force between nucleons is weaker than gravity at short distances; the lower black dashed line is the reference point $G_{s,N} = G_F$ for electroweak-strength couplings. The yellow dashed curve corresponds to a single scattering event per kg-yr exposure in superfluid helium~\cite{Knapen:2016cue}. The pink solid lines are upper bounds on the effective quadratic coupling in the Higgs model of \Eq{eq:g_H} from DM self-interactions ($\phi \phi\to \phi \phi$) and from invisible Higgs decays ($h \to \phi \phi$). The dotted pink line is the structure formation exclusion from Ref.~\cite{Arvanitaki:2014faa}, and the dashed pink line the BBN bound~\cite{Bouley:2022eer} on the nuclear coupling \emph{only} but is less stringent in the quadratic Higgs UV completion. The green line delineates the quadratic nuclear coupling of the QCD axion for $f_a > 10^{8}\,\mathrm{GeV}$.}
\label{fig:G_s_DM}
\end{figure*}

\paragraph*{Models---}
The simplest UV completion for the quadratic scalar coupling of \Eq{eq:G_s_N} is a quadratic coupling to the Higgs doublet $H$:
\begin{alignat}{2}
    \mathcal{L} \supset -\frac{g_H}{2} \phi^2 H^\dagger H, \label{eq:g_H} 
\end{alignat}
with the linear term $\phi H^\dagger H$ forbidden by a $\mathbb{Z}_2$ symmetry, which also guarantees the stability of $\phi$ as a DM candidate (by forbidding decays such as $\phi \to \gamma \gamma$).
At low energies, where the Higgs can be safely integrated out, effective quadratic couplings to nucleons and electrons are generated: $G_{s,N} \simeq (2/9) (g_H m_N / m_h^2 m)$ and $G_{s,e} \simeq (g_H m_e / m_h^2 m)$~\cite{Graham:2015ifn}, where $m_N$, $m_e$, $m_h$ are the nucleon, electron, and Higgs masses. 

The Higgs portal coupling $g_H$ is constrained by the observation that the invisible branching ratio of the Higgs, to which the process $h \to \phi \phi$ contributes, is less than $0.18$~\cite{CMS:2022qva} (falling pink line in \Fig{fig:G_s_DM}). The $g_H$ coupling also leads to quantum corrections to the $\phi$ quartic $\mathcal{L} \supset -\lambda_\phi \phi^4 /4!$ of order $\delta \lambda_\phi \sim g_H^2 / 16\pi^2$, and thus a minimum characteric single-particle self-interaction cross-section $\sigma_\phi \gtrsim \delta \lambda_\phi^2 / (128\pi m^2)$~\cite{Tulin:2017ara}. Astrophysical observations of merging clusters and dwarf galaxies require that $(1 + f_\mathrm{occ}) \sigma_\phi / m \lesssim 1\,\mathrm{cm}^2/\mathrm{g}$~\cite{Randall:2008ppe}, where $f_\mathrm{occ} \simeq (2\pi)^{3/2} n / \sigma_k^3$ is a Bose enhancement factor~\cite{Cruz2024}. (This coherent enhancement of $f_\mathrm{occ}$ has not been discussed in previous literature on self-interacting DM, where the DM constituents are usually assumed to be sufficiently heavy such that $f_\mathrm{occ} \ll 1$.) Self-interactions thus indirectly bound $g_H$ from above as shown by the other pink solid line in \Fig{fig:G_s_DM}, for which I take conservative values of $\rho_\mathrm{DM} = 5 \times 10^6 \, M_\odot / \mathrm{kpc}^3$ and $\sigma_v = 3{,}000 \, \mathrm{km/s}$ for the Bullet Cluster. A conservative estimated bound based on large-scale structure formation for a natural quartic of $|\delta \lambda_\phi| \lesssim 5 \times 10^{-8} (m/\mathrm{eV})^4$~\cite[Sec.~VII]{Arvanitaki:2014faa} is shown as a dotted pink line. Ref.~\cite{Bouley:2022eer} reports Big Bang nucleosynthesis (BBN) constraints on quadratically-coupled dark matter; the corresponding limit on the nuclear coupling is shown by the dashed pink line, but would be dramatically less stringent for the Higgs UV completion, which also includes couplings to electrons and photons which induce a high effective mass for $\phi$ in the early universe. Similarly, $\phi$ receives a mass-squared quantum correction $\delta m^2 \sim g_H \Lambda^2 / 16 \pi^2$ with $\Lambda$ the UV cutoff, which implies that this model has a severe hierarchy problem. As for the Higgs, one may resort to anthropic selection in a landscape of vacua to explain the smallness of $m^2$.

Several of these observational and naturalness constraints can be alleviated (at the expense of introducing others) by introducing a combination of linear and cubic interactions. For example, the super-renormalizable Lagrangian
\begin{alignat}{2}
    \mathcal{L} \supset -A_H \phi H^\dagger H - \frac{A_\phi}{3!} \phi^3 \label{eq:A_H}
\end{alignat}
leads to the same \emph{low-energy} wake force coupling as \Eq{eq:g_H} with the identification $g_H \cong A_H A_\phi / m^2$. Its super-renormalizability means that the model of \Eq{eq:A_H} is UV safe: in particular, the invisible branching ratio of the Higgs is parametrically smaller than for the ``hard'' $g_H$ coupling, and $\phi$ does not suffer from a hierarchy problem. An analogous self-interaction cross-section bound applies to $A_\phi^2$, and there are significant single-exchange (Yukawa) force and stellar cooling constraints on $A_H$. One must also address irreducible cosmological production channels, decays of $\phi$ proportional to $A_H^2$, and the phenomenology of the electroweak phase transition within these classes of models, all left to future work.

Other pathways to generate quadratic interactions of the form $-(G_{s,\psi}m/2)\phi^2 \bar{\psi} \psi$ and $-(G_{p,\psi}m/2)\phi^2 \bar{\psi}i\gamma_5\psi$ are via a light mediator $a$ with mass $m_a$, which couples to $\phi$ with coupling $g_\phi^s$ and to $\psi$ with couplings $g_\psi^s$ and $g_\psi^p$:
\begin{align}
    \mathcal{L} &= \half (\partial \phi)^2 - \frac{m^2}{2} \phi^2 + \half (\partial a)^2 - \frac{m_a^2}{2} a^2 \label{eq:L_light_mediator} \\
    &\phantom{=} - \frac{g_\phi^s}{2} m a \phi^2 - a \bar{\psi} \left(g_\psi^s + g_\psi^p i \gamma_5\right) \psi. \nonumber
\end{align}
(An alternative and more minimal possibility is, as in \Eq{eq:A_H}, through the combination of a cubic self-interaction $\phi^3$ and linear couplings $\phi \bar{\psi} \psi$ and $\phi \bar{\psi} i \gamma_5 \psi$.) Integrating out the mediator $a$, valid for momentum exchange $\max\lbrace|\vect{q}|,|\vect{k}|\rbrace \ll m_a$ yields low-energy quadratic couplings of $G_{s,\psi} \simeq g_\phi^s g_\psi^s / m_a^2$ and $G_{p,\psi} \simeq g_\phi^s g_\psi^p / m_a^2$. A full charting of the parameter space of couplings $g_\phi^s$, $g_\psi^s$, $g_\psi^p$, and masses $m$ and $m_a$ is beyond the scope of this work, but see e.g.~Refs.~\cite{Knapen:2017xzo,Banerjee:2022sqg} in this direction.

Arguably the best-motivated quadratic scalar coupling to nucleons is that of the QCD axion $a$, which irreducibly has (in the parametrization  $-(G_{s,N}/2) m_a a^2 \bar{N} N$ of \Eq{eq:G_s_N}):
\begin{alignat}{2}
    G_{s,N} = \frac{\widetilde{\sigma}}{m_a f_a^2} \quad \text{(QCD axion)},
\end{alignat}
with $\widetilde{\sigma} \approx 15\,\mathrm{MeV}$~\cite{Okawa:2021fto} and $f_a \approx 10^{12}\,\mathrm{GeV}  \, (5.7\,\mathrm{\mu eV}/m_a)$ as a function of the QCD axion mass $m_a$~\cite{GrillidiCortona:2015jxo}. This coupling is depicted by the green line in \Fig{fig:G_s_DM} for $f_a \gtrsim 10^{8}\,\mathrm{GeV}$, the approximate lower bound on the decay constant. 

\paragraph*{Approximate recasted limits---}
Dating back to the experiments of Cavendish, there has been a tremendous effort to search for deviations from Newton's gravitational potential, usually parametrized by the Yukawa potential with strength $\alpha$ (relative to gravity) and range $\lambda$:
\begin{alignat}{2}
    V(r) = - \frac{G_N m_N^2 \alpha}{r} e^{-r/\lambda},
\end{alignat}
as would be generated from a single virtual exchange of a particle with mass $m = 1/\lambda$ and linear coupling $\sqrt{4\pi G_N \alpha}$. The form of the wake potential (\Eq{eq:V_wake_scalar}) is similar at short distances, but the exponential suppression is replaced by the anisotropic form factor $\mathcal{F}(\vect{x})$. 
The \emph{existing} limits on $\alpha^\mathrm{lim}(\lambda)$~\cite{Hoskins:1985tn,Long:1998dk,Chiaverini:2002cb,Hoyle:2004cw,Decca:2005qz,Kapner:2006si,Tu:2007zz,Geraci:2008hb,Yang:2012zzb,Chen:2014oda,Tan:2016vwu,Tan:2020vpf,Lee:2020zjt} can be (approximately) recast into a value of the quadratic coupling $G_{s,N}^\mathrm{lim}$ above which the wake force would have been detectable in past experimental searches---had efforts been made to look for it. I take this recasting map to be:
\begin{alignat}{2}
    G_{s,N}^\mathrm{lim}(m) &=  \min_{r = \lambda} \sqrt{\frac{4 \pi G_N m_p^2 \alpha^{\mathrm{lim}}(\lambda) e^{-r/\lambda}}{\rho_\mathrm{DM} \mathcal{F}_\mathrm{BMB}(r,\theta_r=0)}}. \label{eq:recast_monopole}
\end{alignat}
The $m$ dependence of $G_{s,N}^\mathrm{lim}(m)$ is implicit in the form factor $\mathcal{F}_\mathrm{BMB}$ in the forward direction of the boosted Maxwell-Boltzmann distribution with $\sigma_v = \sigma_k/m \approx 165\,\mathrm{km/second}$ and $v_\mathrm{circ} \simeq \sqrt{2} \sigma_v$, plotted as the red line in the top panel of \Fig{fig:form_radial}. At large $m$ (and fixed $r \sim \lambda$), the form factor decouples as $\mathcal{F} \propto 1/m^2$, so that $G_{s,N}^\mathrm{lim}(m) \propto 1/m$, with the experimental limit from Ref.~\cite{Tan:2020vpf} dominating this tentative ``constraint''. At small $m$, the form factor is constant, so that $G_{s,N}^\mathrm{lim}$ tends to a constant. In the latter regime, the wake potential scales as $1/r$ and is only distinguishable from gravity via its generic violation of the weak equivalence principle (i.e.~dependence on chemical composition) and the apparent mismatch of the effective value of $G_N$ at short range and on astronomical scales.

The map of \Eq{eq:recast_monopole}, shown as the blue filled region in \Fig{fig:G_s_DM}, cannot be deemed a limit with a quantifiable confidence level, because the geometry and orientation of the experiment relative to the DM wind must be taken into account. The experiments with leading limits on $\alpha(\lambda)$ did not explicitly search for composition dependence (only deviations from the inverse-square-radius behavior of the force), although they can be (and have been) modified to do so \cite{Wagner:2012ui}. Nevertheless, the recasting via \Eq{eq:recast_monopole} should be an accurate (if conservative) guide to the approximate sensitivity of existing experimental setups.

\paragraph*{Sensitivity projections---}
The sensitivity of dedicated wake force experiments can be estimated faithfully. As a representative case, consider the wake force exerted on a ``target'' tungsten sphere of radius $R = 1\,\mathrm{cm}$ and mass density $\rho_N = 19.28 \, \mathrm{g/cm^3}$ positioned in the forward direction of the wake emanating from an identical ``source'' sphere, with a center-to-center separation of $2.2 R$. This arrangement should be close to the optimal geometry given a target of size $R$: making the source much larger than the target does not parametrically increase the signal in the regime of large $m \sigma_k R$ in view of the large-radius scaling of the wake force $|\vect{F}| \propto \mathcal{F}_\mathrm{BMB} / r^2 \propto 1/r^4$. For this geometry, the wake force on the target points towards the source, and has a parametric size:
\begin{alignat}{2}
    |\vect{F}| \sim \frac{G_{s,N}^2 \rho_\mathrm{DM}}{4\pi R^2} \left(\frac{M}{m_N}\right)^2 \frac{1}{1+(m \sigma_v R)^2} \label{eq:F_wake_parametric},
\end{alignat}
with $M = (4\pi/3) \rho_N R^3 \approx 81\,\mathrm{g}$ the mass of the spheres.

Suppose the target sphere is confined to its position in a (generalized) harmonic oscillator potential well of angular frequency $\omega = 2\pi \,\mathrm{mHz}$ and losses corresponding to a quality factor $Q = 10^8$, as one may obtain with a future torsion-balance (TB) setup~\cite{Arvanitaki:2023fij}. In this rather generic case, the minimal variance on the force measurement after an integration time $t_\mathrm{int} = 3\,\mathrm{yr}$ is given by the fluctuation-dissipation theorem~\cite{Kubo:1966}:
\begin{alignat}{2}
    \sigma_F^2 = \frac{2 \mathcal{E}(\omega,T) \omega M}{Q t_\mathrm{int}}. \label{eq:variance_F}
\end{alignat}
In the above, $\mathcal{E}(\omega,T)$ is the noise temperature, which is typically (at least) the thermodynamic temperature $T$ of the environment; in thermal equilibrium, $\mathcal{E}(\omega,T) = \omega \left[1/2 + 1/(e^{\omega/T}-1) \right]$~\cite[Ch.~17.2]{Mandel:1995}. With active cooling methods and a sufficiently decoupled oscillator, one could in principle achieve the standard quantum limit (SQL) of $\mathcal{E}(\omega,T) = \omega / 2 \approx 2 \times 10^{-14} \, \mathrm{K}$, though present-day capabilities are far from these values. \Eq{eq:variance_F} represents the minimum contribution from thermal/quantum noise, and does not include readout/back-action and environmental noise sources.

This corresponding idealized coupling sensitivity, at unit signal-to-noise ratio $|\vect{F}|/\sigma_F = 1$, of this representative oscillator is depicted in \Fig{fig:G_s_DM} as blue lines, at the thermal noise limit $\mathcal{E} = T = 10\,\mathrm{mK}$ (solid) and at the SQL $\mathcal{E} = \omega/2$ (dashed). For this forecast, the size of the wake force between the spheres is computed numerically, as opposed to the parametric estimate of \Eq{eq:F_wake_parametric}, and the target-source separation is assumed to remain aligned with $\vect{v}_\mathrm{circ}$ of \Eq{eq:mom_BMB}. Any realized sensitivity will be degraded by an order-unity factor based on the desired discovery or exclusion threshold, and for a non-corotating experimental setup.Especially the thermal-noise-limited sensitivity is a benchmark that current experimental setups may strive towards; attaining it would have significant discovery potential for the DM wake force of e.g.~the quadratic Higgs coupling in \Eq{eq:g_H}. Dedicated wake force experiments may mitigate systematics through various control knobs, such as its weak-equivalence-principle-violating nature, distance dependence, and most strikingly, its asphericity and angular variation with respect to the DM wind direction, which changes on a diurnal basis in the lab frame. It would be interesting to explore the optimal experimental geometry and orientation to maximize the wake force signal and minimize more prosaic backgrounds. For example, short-range \emph{anisotropic} corrections to gravity can be sensed by modified torsion-balance setups~\cite{Bobowski:2022qqa}.

\paragraph*{Other DM experiments---}
Traditional DM experiments searching for isolated scattering events (which produce low-energy quasiparticles in the scattering target) lose sensitivity at low DM masses, where the recoil energy is suppressed. The effective coupling at which one expects a single scattering event for a kg-yr exposure of the low-threshold superfluid helium proposal from Refs.~\cite{Schutz:2016tid,Knapen:2016cue} is shown in \Fig{fig:G_s_DM} as the dashed yellow line. Wake force experiments instead become \emph{more} sensitive at low masses due to the increased range, and may thus probe effective single-particle cross-sections far below $10^{-40} \, \mathrm{cm}^2$ and even $10^{-50}\, \mathrm{cm}^2$ (gray dashed lines). In this sense, DM wake force searches can bridge the gap between traditional DM scattering experiments and coherent interactions of bosonic DM with macroscopic targets (haloscopes).

\subsection{Neutrino Detection} \label{sec:neutrino}

\paragraph*{Cosmic neutrino background (C$\nu$B)---} 
The present-day C$\nu$B is, in a standard cosmology, expected to have a temperature: 
\begin{align}
    T_\nu \approx 1.95\,\mathrm{K} \approx 1.68 \times 10^{-4} \, \mathrm{eV} \approx \frac{1}{1.17 \times 10^{-3} \,\mathrm{m}},
\end{align}
which also sets the characteristic momentum and the corresponding coherence length $T_{\nu}^{-1}$ of the neutrino background (and wake force). The number densities per mass eigenstate are $n_{\nu_i} = \bar{n}_{\nu_i} \approx 56 \, \mathrm{cm}^{-3}$.
Given the observed mass-squared splittings of the neutrino mass eigenstates, at least two (and possibly all three) are nonrelativistic. As shown in \App{app:fermions} and in Refs.~\cite{Horowitz:1993kw,Ferrer:1998ju,Ferrer:1999ad,Ghosh:2022nzo, Arvanitaki:2022oby,Arvanitaki:2023fij,Blas:2022ovz}, wake forces can be mediated by fermions too.

In \App{app:four_fermi}, I calculate the full wake potential for a single Dirac-neutrino mass eigenstate interacting through neutral-current interactions with SM fermions $\psi$ with vector and axial-vector couplings $g_\psi^V$ and $g_\psi^A$, respectively. Those results are easily generalized to include three Dirac neutrinos and the charged-current interactions. However, the wake potential is largest for the highest-mass neutrino eigenstate, so in the ``normal'' mass hierarchy case with strong ordering $m_1 \lesssim m_2 \ll m_3$, it is a fine approximation up to $\mathcal{O}(m_2/m_3)$ corrections to take into account only the heaviest neutrino $\nu_3$. Neutrino \emph{oscillations} from the charged-current interactions can similarly be neglected, as the mass-squared splittings between the neutrinos are higher than the square of the neutrino temperature: $|\Delta^2| \gg T^2$ in the language of the discussion around \Eqs{eq:Delta}--\ref{eq:osc}.

With those assumptions, the monopole-monopole wake potential mediated by the C$\nu$B between two SM fermions $\psi_1$ and $\psi_2$ can be translated from \Eq{eq:V_Dirac_VV_2} (for Dirac neutrinos) to:
\begin{alignat}{2}
    V(\vect{x}) = - \frac{2 G_F^2 Q^{3,V}_1 Q^{3,V}_2 m_3 (n_{\nu_3}+\bar{n}_{\nu_3})}{4\pi r} \mathcal{F}_{VV}(\vect{x}), \label{eq:V_CNB_VV}
\end{alignat}
where $Q^{3,V}_\psi = g_\psi^V - U_{3e} U_{e3}^\dagger \delta_{\psi e}$ is the effective vector ``charge'' of the fermion $\psi$ as seen by $\nu_3$, including both the neutral-current vector coupling $g_\psi^V$ and the charged-current vector coupling $U_{3e} U_{e3}^\dagger$ (only to electrons). The form factor $\mathcal{F}_{VV}(\vect{x})$ is given in \Eq{eq:form_Dirac_VV} (where one should take $m = m_3$), which in the nonrelativistic limit matches \Eq{eq:form_general} up to $\mathcal{O}(T^2/m^2)$ corrections and thus that of the top panel in \Fig{fig:wake_BFD} for a BFD distribution.

The characteristic strength of the C$\nu$B wake force can be expressed in terms of its relative size $\alpha_\mathrm{C\nu B}$ relative to the gravitational force between two macroscopic bodies $1$ and $2$:
\begin{alignat}{2}
    \alpha_\mathrm{C\nu B} &\simeq \frac{2 G_F^2 \big\langle Q^{3,V}_1 \big\rangle \big\langle Q^{3,V}_2\big\rangle m_3 (n_{\nu_3}+\bar{n}_{\nu_3})}{4\pi G_N m_N^2} \label{eq:alpha_nu_1} \\
    & \approx 1.9 \times 10^{-22} \, \big\langle Q^{3,V}_1 \big\rangle \big\langle Q^{3,V}_2\big\rangle \left(\frac{m_3}{60\,\mathrm{meV}}\right) \label{eq:alpha_nu_2}
\end{alignat}
at short distances $r \ll T_\nu^{-1} \approx 1.17\,\mathrm{mm}$, where $\big\langle Q^{3,V}_i\big\rangle$ is the average effective vector charge per nucleon of the heaviest neutrino mass eigenstate in the two bodies $i = 1,2$. The present best constraint on $\alpha(T_\nu^{-1})$ is about $10^{-3}$~\cite{Yang:2012zzb}, far worse than \Eq{eq:alpha_nu_2}. An improvement by 19 orders of magnitude is not in reach of any conceivable force experiment; even the SQL-limited sensitivity of the dashed line in \Fig{fig:G_s_DM} would ``only'' constitute a $10^{13}$ improvement in $\alpha(T_\nu^{-1})$.

The monopole-dipole wake potential mediated by the C$\nu$B between two SM fermions $\psi_1$ and $\psi_2$ can similarly be translated from \Eqs{eq:V_Dirac_VA} and \ref{eq:form_Dirac_VA}. The precise expression of the form factor is not important for a back-of-the-envelope estimate of its size, which is $\mathcal{O}(T/m)$. For a spherical tungsten source mass $1$ of radius $R \sim T^{-1}$ (larger sources are not relevant given the short range), the integrated monopole-dipole wake potential for a polarized target with (nuclear or electronic) spin $\vect{\sigma}_2$ is of the parametric form $V(\vect{\sigma}_2) \sim \vect{\Omega} \cdot \vect{\sigma}_2$, with an angular precession frequency:
\begin{alignat}{2}
    |\vect{\Omega}| &\sim \frac{G_F^2 (n_\nu+\bar{n}_{\nu})T^2}{4\pi} \frac{4\pi \rho_N}{3 m_N T^3} \sim 10^{-28} \, \mathrm{rad/s}.\label{eq:Omega_CNB_2} 
\end{alignat}
In comparison, the C$\nu$B wind effect linear in $G_F$ from Ref.~\cite{Stodolsky:1974aq} is of the form 
\begin{align}
    V(\vect{\sigma}) &\sim G_F (n_\nu - \bar{n}_\nu) \langle  \vect{v}_\nu \rangle \cdot \vect{\sigma} \label{eq:Omega_CNB_3}\\
    &\sim 10^{-23} \, \mathrm{rad/s}, \langle  \vect{v}_\nu \rangle \cdot \vect{\sigma} \frac{n_\nu - \bar{n}_\nu}{n_\nu + \bar{n}_\nu}. \label{eq:Omega_CNB_4}
\end{align}
The precession induced from the wake potential in \Eq{eq:Omega_CNB_2} is larger than that of \Eq{eq:Omega_CNB_4} if the neutrino asymmetry is as small as the baryon asymmetry, i.e.~$n_\nu - \bar{n}_\nu \sim 10^{-10} (n_\nu + n_{\bar{\nu}})$. For neutrino asymmetries larger than about $10^{-5}$, the effect from \Eq{eq:Omega_CNB_4} would exceed that of the C$\nu$B wake potential in any geometry. However, either effect appears to be far below the quantum projection noise limits for nuclear and electron spins~\cite{Budker:2006gya}.

\paragraph*{Solar and reactor neutrinos---}
There exist more intense sources of \emph{relativistic} neutrinos, for which the heuristic form of the wake potential in \Eq{eq:V_heuristic} still holds, now with $\rho \simeq |\vect{k}_0| n$ instead of $\rho \simeq m n$, as can be seen from taking the relativistic limits of \Eqs{eq:form_Dirac_VV} and~\ref{eq:form_Dirac_VA}. The solar neutrino flux and thus number density is about $n_{\nu, \odot} \approx 7 \times 10^{10}~\mathrm{cm}^{-2}\,\mathrm{s}^{-1} \approx 2 \, \mathrm{cm}^{-3}$, but with typical energies of $E_{k_0} \sim \mathrm{MeV}$. 
Solar neutrinos lead to a fractional correction to gravity on the order of $\alpha_{\nu,\odot} \sim 2 G_F^2 E_{k_0} n_{\odot,\nu}  / (4 \pi G_N m_N^2) \sim 10^{-16}$ at distances of $k_0^{-1} \sim 10^{-13}\,\mathrm{m}$, again far too small to be detected, even after taking into account the neutrino oscillation effects described in \Sec{sec:range}. Fluxes from reactor neutrinos, whose individual coherent scattering events have been detected already~\cite{COHERENT:2017ipa}, can be a couple of orders of magnitude larger, but not enough to close the gap towards practical observability of neutrino wake forces with present-day technology. (The same conclusion holds for neutrinos from other sources, such as supernovae, radioactive samples, and spallation.)

\section{Comparisons} \label{sec:comparisons}
In this section, I compare the size of wake potential effects to other phenomena that necessarily occur in the same theories with quadratic interactions to matter. Other than in \Sec{sec:double_exchange}, the focus will be on the interaction of a light field $\phi$ coupled to nucleons as in \Eq{eq:G_s_N}, since some comparisons for neutrino wake forces have already been made in \Sec{sec:neutrino}.

\subsection{Double-Exchange Forces} \label{sec:double_exchange}

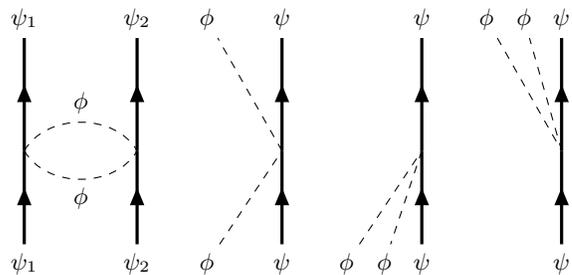
\begin{figure}[t]
    \begin{tikzpicture}
      \begin{feynman}
        \vertex (i1) {\(\psi_1\)};
        \vertex [above=of i1] (v1);
        \vertex [above=of v1] (f1) {\(\psi_1\)};
        \vertex [right=of i1] (i2) {\(\psi_2\)};
        \vertex [above=of i2] (v2);
        \vertex [above=of v2] (f2) {\(\psi_2\)};
        \diagram* {
        (i1) -- [fermion, very thick] (v1) -- [fermion, very thick] (f1),
        (i2) -- [fermion, very thick] (v2) -- [fermion, very thick] (f2),
        v1 -- [scalar, edge label=\(\phi\), loop, in=120, out=60, looseness=1] v2, 
        v1 -- [scalar, edge label'=\(\phi\), loop, in=240, out=300, looseness=1] v2, 
        };
      \end{feynman}
    \end{tikzpicture} 
    \quad 
    \begin{tikzpicture}
      \begin{feynman}
        \vertex (i1) {\(\psi\)};
        \vertex [above=of i1] (v1);
        \vertex [above=of v1] (f1) {\(\psi\)};
        \vertex [left=1cm of i1] (i2) {\(\phi\)};
        \vertex [left=1cm of f1] (f2) {\(\phi\)};
        \diagram* {
        (i1) -- [fermion, very thick] (v1) -- [fermion, very thick] (f1),
        (i2) -- [scalar] (v1) -- [scalar] (f2),
        };
      \end{feynman}
    \end{tikzpicture} 
    \quad 
    \begin{tikzpicture}
      \begin{feynman}
        \vertex (i1) {\(\psi\)};
        \vertex [above=of i1] (v1);
        \vertex [above=of v1] (f1) {\(\psi\)};
        \vertex [left=1cm of i1] (i2) {\(\phi\)};
        \vertex [left=0.5cm of i1] (f2) {\(\phi\)};
        \diagram* {
        (i1) -- [fermion, very thick] (v1) -- [fermion, very thick] (f1),
        (i2) -- [scalar] (v1) -- [scalar] (f2),
        };
      \end{feynman}
    \end{tikzpicture} 
    \quad
    \begin{tikzpicture}
      \begin{feynman}
        \vertex (i1) {\(\psi\)};
        \vertex [above=of i1] (v1);
        \vertex [above=of v1] (f1) {\(\psi\)};
        \vertex [left=1cm of f1] (i2) {\(\phi\)};
        \vertex [left=0.5cm of f1] (f2) {\(\phi\)};
        \diagram* {
        (i1) -- [fermion, very thick] (v1) -- [fermion, very thick] (f1),
        (i2) -- [scalar] (v1) -- [scalar] (f2),
        };
      \end{feynman}
    \end{tikzpicture} 
    \caption{Diagrams for double-exchange force (left), coherent scattering (second from left), and in-medium potentials (right three diagrams).} \label{fig:diagram_other}
\end{figure}

The diagrams in \Fig{fig:feyn_scalar} can be regarded as ``cut'' versions of the two-particle-exchange diagram in \Fig{fig:diagram_other}, with the $\phi$ propagator split into two external legs which have nonzero occupation number. The resulting wake potential is a purely classical wave effect, as evident from \Sec{sec:classical}. The double-exchange potential from \Fig{fig:diagram_other} is a quantum effect arising from vacuum fluctuations, as can be shown by loop- or $\hbar$-counting arguments. For the coupling in \Eq{eq:L_scalar}, the double-scalar-exchange potential between two $\psi$ particles is~\cite{Ferrer:1998ue,Ferrer:2000hm,Bauer:2023czj}:
\begin{alignat}{2}
    V_{2 \phi} &= -\frac{G^2 m^3}{128 \pi^3 r^2} K_1(2 m r), \label{eq:V_2phi} 
\end{alignat}
where $K_1$ is a modified Bessel function of the second kind. 
Firstly, the range of the force from the exchange of two virtual scalars is roughly half the Compton wavelength of $\phi$, whereas wake forces have a range of order the de Broglie wavelength of the ambient $\phi$ particles. Secondly, for a force experiment carried out at $r = 1/2m$, the ratio of the double-exchange and wake potentials is:
\begin{align}
\frac{V_\mathrm{wake}}{V_{2 \phi}} = \frac{16 \pi^2}{K_1(1)} \frac{n}{m^3} \quad \left( r = \frac{1}{2m} \right).
\end{align}
The wake force is enhanced at large number densities, and, unlike the double-exchange force, is not suppressed by a loop factor. A range of $1\,\mathrm{mm}$ or above, where force experiments achieve their best sensitivity, corresponds to $m \lesssim 1\,\mathrm{meV}$ to avoid exponential suppression in the double-exchange force. If $\phi$ makes up all of the DM, then this implies $n/m^3 \gtrsim 10^6$ and thus a much larger wake force than double-exchange force. The occupation number decreases quickly for sub-millimeter Compton wavelengths, but so do the sensitivities of force experiments, so that the longer range of the wake force by a factor of $\mathcal{O}(\sigma_v^{-1})$ is a major advantage. Thus, if $\phi$ is a nonnegligible fraction of DM, then its wake force is always the dominant effect in force experiments, and the double-exchange force is negligible. (I do not attempt to recast $\alpha^\mathrm{lim}(\lambda)$ in terms of the double-exchange force sensitivity to $G_{s,N}$, because current experimental setups are not sensitive to couplings in the perturbative, unscreened regime below the brown region in \Fig{fig:G_s_DM}.)

Similar arguments apply to double-fermion-exchange potentials. For SM neutrinos at distances $r \ll (2 m_\nu)^{-1}$, they are of the form $V_{2\nu} \sim G_F^2 / (4\pi^3 r^5)$ and contain both spin-independent and spin-dependent (dipole-dipole) components \cite{Feinberg:1968zz,Feinberg:1989ps,Hsu:1992tg}. This phenomenon is far too small to be observable with current technology, except possibly at very short distances with future experimental and theoretical efforts in muonium spectroscopy~\cite{Dzuba:2017cas}. For fermionic DM, one must have $n/m^3 \lesssim \sigma_v^3$ by the Pauli exclusion principle, which is the primary reason for why fermionic DM wake forces are harder to detect. 

\subsection{Elastic scattering} \label{sec:elastic}
For $m \ll m_\psi$ and for momentum transfers where the nonrenormalizable operator of \Eq{eq:L_scalar} is valid, $\phi$ particles can scatter elastically off SM $\psi$ particles, with a cross-section:
\begin{align}
\sigma_\mathrm{el} \simeq \frac{G^2 m^2}{4\pi}. \label{eq:xsec_elastic}
\end{align}
In the limit of $m \gg m_\psi$ or where the momentum transfer in the interaction is above the effective cutoff of the interaction (as would be possible in e.g.~the light-mediator UV completion of \Eq{eq:L_light_mediator}), there are generally further suppression factors, which only strengthen the point below.

\emph{Individual} scattering events occur stochastically, and are in practice only measurable if suitable quasiparticles with sufficiently low energy threshold can be excited and detected. The ultimate reach for single-phonon excitation in superfluid helium~\cite{Schutz:2016tid,Knapen:2016cue}, which incurs additional suppression factors beyond \Eq{eq:xsec_elastic}, is shown as the dashed yellow line in \Fig{fig:G_s_DM}. 
An alternative approach is to look for the collective, integrated recoil of many elastic scattering events on a macroscopic target over time; each scattering event will, on average, impart some momentum to the target in the direction of the DM wind~\cite{Day:2023mkb}.

In a bath of nonrelativistic $\phi$ particles with number density $n$ and typical momentum magnitude $|\vect{k}_0|$, the scattering rate per $\psi$ particle is $\Gamma \sim n \sigma_\mathrm{el} |\vect{k}_0|/m$. If the average momentum transfer per scattering event is $\overline{\vect{k}}_0$, the time-averaged force due to elastic scattering, on a single $\psi$ particle, is:
\begin{align}
\vect{F}_\text{el} \sim \frac{G^2 n m}{4\pi} |\vect{k}_0| \overline{\vect{k}}_0. \label{eq:F_el_scalar}
\end{align}
The wake force $\vect{F} = -\vect{\nabla} V(\vect{x})$ induced by the potential in \Eq{eq:V_wake_scalar} with e.g.~the MB form factor of \Eq{eq:form_MB} (for simplicity) is equal to:
\begin{align}
\vect{F} = -\hat{\vect{x}} \frac{G^2 n m}{4\pi} \sigma_k^2 e^{-2\sigma_k^2 r^2}  \left(4+\frac{1}{\sigma_k^2 r^2} \right).
\label{eq:F_wake_scalar_MB}
\end{align}
Comparison of \Eqs{eq:F_el_scalar} \&~\ref{eq:F_wake_scalar_MB} shows that the elastic scattering and wake forces on a single $\psi$ particle are roughly equal in magnitude when the two $\psi$ particles separated by a distance $r \sim 1/\sigma_k$ and if $|\vect{k}_0| |\overline{\vect{k}}_0| \sim \sigma_k^2$. However, the wake force is larger than the elastic scattering force for $r \lesssim 1/\sigma_k$ or if the average momentum transfer $|\overline{k}_0|$ is much smaller than the typical momentum, which is the case for the C$\nu$B and likely other types of primordial dark radiation.

To maximize detection prospects, one generally wants to measure either force on as many $\psi$ particles as possible. The elastic scattering force is coherent when the target size is smaller than the coherence length of $\phi$, $R_\text{target} \lesssim 1/\sigma_k$, leading to a scaling of $(n_{\psi,\text{target}} R_\text{target}^3)^2$ with the number of particles in that regime. The wake force enjoys a similar coherent enhancement as $(n_{\psi,\text{target}} R_\text{target}^3)\times(n_{\psi,\text{source}} R_\text{source}^3)$ for $R_\text{target}, R_\text{source} \lesssim 1/\sigma_k$. However, in certain cases of practical relevance (e.g.~torsion balances, optically-levitated dielectrics, atom interferometry), maximum acceleration sensitivity may be achieved on a very small target. Having the option to use a larger \emph{source} of $\psi$ particles over which the wake force is coherent can be advantageous in such cases. Furthermore, the wake force is generally easier to control and manipulate, as it depends on the density, size, distance, and (generally) chemical composition of the source, allowing for a variety of experimental knobs to be turned. Both coherent elastic scattering forces and wake forces enjoy a cosmic reference direction, which should help mitigate systematics. However, like any pair-wise force, wake forces can be made to vary temporally---by employing rotary stages~\cite{Tan:2020vpf,Lee:2020zjt,Hoyle:2004cw,Kapner:2006si,Chen:2014oda,Tan:2016vwu}, driven mechanical oscillators~\cite{Long:1998dk}, or linear motions of source~\cite{Chiaverini:2002cb,Hoskins:1985tn,Tu:2007zz,Geraci:2008hb,Yang:2012zzb} or target masses~\cite{Decca:2005qz}---allowing them to be probed by AC experiments as opposed to the daunting experimental challenge of searching for an uncontrollable DC force. 
Ref.~\cite{Day:2023mkb} computes elastic scattering and the associated DC forces in the regime where screening is important, and discusses detection prospects of these forces with satellite tests of the equivalence principle and torsion balance experiments. 

\subsection{In-Medium Potentials and Forces} \label{sec:inmedium}

In a sea of $\phi$ particles, individual $\psi$ particles experience an in-medium potential and force at \emph{linear} order in $G$, represented by the right three diagrams of \Fig{fig:diagram_other}, independent of any neighboring $\psi$ particles. For the quadratic scalar field interaction of \Eq{eq:L_scalar}, this in-medium potential is simply:
\begin{align}
V_\text{IM} = \frac{G m}{2} \phi^2 \simeq \frac{G n}{2} \left[1+\cos(2\omega t-2\vect{k}\cdot \vect{x}) \right], \label{eq:V_IM_scalar}
\end{align}
with the second equality for a simple plane wave of $\phi$, for which the in-medium force is:
\begin{align}
\vect{F}_\text{IM}= G n \vect{k} \sin(2\omega t-2\vect{k}\cdot \vect{x}). \label{eq:F_IM_scalar}
\end{align}
Diagrammatically, the ``constant'' term in \Eq{eq:V_IM_scalar} arises from the same diagram as elastic scattering (second from left in \Fig{fig:diagram_other}, with both positive- and negative frequency factors), whereas the oscillatory terms at $2\omega$ in \Eq{eq:V_IM_scalar} and the force of \Eq{eq:F_IM_scalar} arise from the rightmost two diagrams in \Fig{fig:diagram_other}. For a random superposition of waves, $n$ itself is stochastically varying with coherence times and lengths of order $m / \sigma_k^2$ and $1/\sigma_k$, respectively, generating additional forces from spatial gradients in $n$, which are of the same order of magnitude as the oscillatory force in \Eq{eq:F_IM_scalar}.

In the perturbative regime for $G$, these in-medium potentials and forces are naively larger than their corresponding wake counterparts. In particular, $V_\text{wake} / V_\text{IM}  \sim  G m / 2\pi r \lesssim 1$ for a single source particle due to perturbativity; for many source particles over a de Broglie volume of $\mathcal{O}(k_0^{-3})$, $V_\text{wake} / V_\text{IM} \sim G m n_\psi / k_0^2 \lesssim 1$ for a cosmic field when screening effects are negligible (\Sec{sec:screening}). However, in the ``high-mass'' regime of \Fig{fig:G_s_DM}, the fractional nucleon mass variation is $V_\mathrm{IM} / m_N \sim G_{s,N} n/ (2 m_N) \sim 10^{-33} (G_{s,N}/\mathrm{GeV}^{-2}) (\mathrm{eV}/m)$, well out of reach of atomic-clock sensitivity or that of other metrological experiments. The force from \Eq{eq:F_IM_scalar} oscillates at a frequency of $m/2\pi \sim 242 \, \mathrm{THz} \, (m/\mathrm{eV})$, too rapidly to be detected by torsion balances or other force sensors over the mass range in \Fig{fig:G_s_DM}. Similarly, the stochastic acceleration from gradients in $n$ is about $10^{-12} \, \mathrm{m/s^2} (G_{s,N}/\mathrm{GeV}^{-2})$ independent of mass, but also has a prohibitively short coherence time of $2\pi / (m \sigma_v^2 / 2) \approx 273\,\mathrm{\mu s} \, (\mathrm{eV}/m)$, after which it averages to zero.

The considerations for in-medium potentials from fermions are similar; the effect from Ref.~\cite{Stodolsky:1974aq} and \Eq{eq:Omega_CNB_3} is one such spin-dependent incarnation for the C$\nu$B, and the only known nonstochastic effect linear in $G_F$ for the C$\nu$B~\cite{Langacker:1982ih}.

\subsection{Screening} \label{sec:screening}
In this work, I have so far treated the quadratic $\phi$ interactions perturbatively, in which the separation between a cosmic background wave and a linear superposition of scattered waves from all perturbing sources is justified. However, at large $G$ and/or small momentum, this perturbative expansion can break down~\cite{Olive:2007aj,Hees:2018fpg,Arvanitaki:2022oby}. For an arbitrary but static source distribution with number density $n_\psi(\vect{x})$, the full $\phi$ field obeys the Klein-Gordon equation
\begin{align}
\big[\Box + m^2 + G m n_\psi(\vect{x})\big]\phi(\vect{x}) = 0
\end{align}
with a spatially varying effective mass-squared $m_\mathrm{eff}^2 = m^2 + G m n_\psi(\vect{x})$, which one can model as a corresponding index of refraction~\cite{Arvanitaki:2022oby,Arvanitaki:2023fij}. 

To illustrate the breakdown of the perturbative treatment at large $G$ and/or small momenta, consider the 1D problem in the $z$-direction with a step function in number density, which vanishes for $z<0$ and is a constant $n_\psi$ for $z>0$. Suppose there is an incoming wave of the form $\cos(\omega t - k_0 z)$ for $z < 0$. A wave will be reflected from the step at $z = 0$: $R_c \cos(\omega t + k_0 z) + R_s \sin(\omega t +  k_0 z)$. Likewise, a wave is transmitted inside the medium ($z>0$) with wavenumber:
\begin{align}
k_\psi = \sqrt{k_0^2 - G m n_\psi}.
\end{align}
For real $k_\psi$, write the transmitted solution as $T_c^+ \cos(\omega t - k_\psi z) + T_s^+ \sin(\omega t - k_\psi z)$, and for imaginary $k_\psi$, as $e^{-|k_\psi|z} [T^-_c \cos(\omega t) + T^-_s \sin(\omega t)]$. This simple boundary problem can be solved for the reflection and transmission coefficients:
\begin{alignat}{4}
    T^+_c &= \frac{2 k_0}{k_0 + k_\psi}, \quad &&R^+_c = \frac{k_0 -k_\psi}{k_0 + k_\psi}, \\
    T^-_c &= \frac{2 k_0^2}{k_0^2 + |k_\psi|^2}, \quad &&R^-_c = \frac{k_0^2 - |k_\psi|^2}{k_0^2 + |k_\psi|^2},\\
    R^+_s &= T^+_s = 0, \quad &&R^-_s = T^-_s = \frac{-2k_0 |k_\psi|}{k_0^2 + |k_\psi|^2}.
\end{alignat}
for real (${}^+$) and imaginary (${}^-$) $k_\psi$, respectively. The fraction of transmitted power is $4 k_0 k_\pi / (k_0 + k_\psi)^2$ for $G m n_\psi < k_0^2$ and vanishes for $G m n_\psi > k_0^2$. Either way, most or all of the incoming wave is reflected when $|G| m n_\psi >  k_0^2$. (Disastrous effects due to tachyonic instabilities could theoretically occur for large negative $G$, which I will not consider here.)

This ``screening'' effect has important implications for detection of the wake force from a cosmic field. For sufficiently large coupling, most of the ``dark'' field could be reflected from the environment, and not make it to the laboratory setup.
In general, the perturbative expansion can be expected to hold at large distances if
\begin{align}
|G| m n_\psi \ll  \sigma_k^2, \label{eq:no_screening}
\end{align}
where $n_\psi$ is the typical density of the environment (Earth's crust or atmosphere, laboratory walls). In \Fig{fig:G_s_DM}, the regime where \Eq{eq:no_screening} is \emph{not} satisfied is shaded in light brown, for $n_\psi$ corresponding to the nucleon density of aluminum.
Even when \Eq{eq:no_screening} is satisfied, \emph{some small fraction} of incoming waves will be shielded by the environment~\cite{Arvanitaki:2022oby}, but when the inequality is strong, the perturbative expansion of this work should be valid to a good approximation.

A full calculation of screening effects would be highly specific to the geometry of the system under study, and is left to future work, though the approach of Ref.~\cite{Hees:2018fpg} should be valid for DM masses $m \ll 10^{-11}\,\mathrm{eV}$, i.e.~for vacuum de Broglie wavelengths larger than Earth's radius. In any case, given self-interaction and structure formation bounds on the model space discussed in \Sec{sec:dmsearches}, there is a significant theoretical bias (for technically natural self-couplings) towards low values of $|G|$ that satisfy \Eq{eq:no_screening}. Close to the boundary of \Eq{eq:no_screening}, the calculation of the wake force to $\mathcal{O}(G^3)$ may be merited. Based on the simulations of \App{app:nonperturbative}, I speculate that the next-to-leading order wake potential will generate a fractional correction to the leading $\mathcal{O}(G^2)$ wake force of $\mathcal{O}(G m n_\psi/k_0^2)$. Shielding and screening effects are already negligible at the leading edge of present-day sensitivity (lower boundary of blue region in \Fig{fig:G_s_DM}), and will be even more so for future (wake) force experiments.

\section{Discussion} \label{sec:discussion}
In this work, I have introduced a general formalism for the perturbative calculation of wake forces and potentials, which can be understood as classical wave phenomena that generically occur in a ``sea'' of particles quadratically coupled to nonrelativistic source and target particles. I have computed the analytic form of wake forces from a variety of quadratic interactions of both (pseudo)scalars (\Sec{sec:classical}, \Sec{sec:quantum}, and \App{app:scalars}) and fermions (\App{app:fermions}), as well as their spatial profiles and range (\Sec{sec:range}). Before outlining potential future applications and directions, I will comment on the overlap and differences with previous literature (of which I am aware).

Firstly, Ref.~\cite{Hees:2018fpg} is a study of both linearly- and quadratically-coupled light scalar DM, wherein the in-medium potentials of \Sec{sec:inmedium} and some screening phenomena (\Sec{sec:screening}) were discussed for spherical sources (Earth in particular). The screening effects are nothing but the nonperturbative manifestation of the wake potential. The approach of Ref.~\cite{Hees:2018fpg} is valid in the regime where the de Broglie wavelength of the mediating particles is much larger than the size of the source, but finite-wavenumber effects and thus the short-range nature of the wake force were neglected, so their constraints should be quantitatively reconsidered for $m \gtrsim 10^{-11}\,\mathrm{eV}$. Separately, Ref.~\cite{Berezhiani:2018oxf} found an attractive $1/r^2$ force on galactic scales between test particles in a Bose-Einstein condensate (BEC) of quadratically-coupled scalar field DM with self-interactions. In the limit of vanishing self-interactions, their results are equivalent to the perturbative wake force formalism of \Sec{sec:classical} at zero background momentum.

Wake forces in a neutrino medium have also been posited in Refs.~\cite{Horowitz:1993kw,Ferrer:1998ju,Ferrer:1999ad,Ghosh:2022nzo, Arvanitaki:2022oby,Arvanitaki:2023fij}, and in a bosonic medium in Refs.~\cite{Ferrer:2000hm,Arvanitaki:2023fij}, mostly independent of this work. Refs.~\cite{Horowitz:1993kw,Ferrer:2000hm,Ghosh:2022nzo} regarded the medium-dependent forces as a version of the two-particle-exchange diagram, with the effective propagators modified by the (thermal) background.  Refs.~\cite{Horowitz:1993kw,Ghosh:2022nzo} also found an additional $1/(Tr)^4$ suppression of the $G_F^2 \rho_\mathrm{C\nu B}/r$ potential at large distances for an isotropic neutrino background. Ref.~\cite{Ferrer:2000hm} primarily discussed the case of a Bose-Einstein condensate background. The background-modified propagator approach of Refs.~\cite{Horowitz:1993kw,Ferrer:1998ju,Ferrer:1999ad,Ghosh:2022nzo} is equivalent to my approach, though the kinetic-theory formalism of \Sec{sec:quantum} is more efficient for wake force calculations for arbitrary, anisotropic phase-space distributions. (\Eqs{eq:V_Dirac_VV_2},~\ref{eq:form_Dirac_VV},~and~\ref{eq:form_FD} match Eqs.~3.24~and~3.25 of Ref.~\cite{Ghosh:2022nzo} with $n_\nu = n_{\bar{\nu}} = 3 \zeta(3) T_\nu^3 / 4 \pi^2$ per generation.) 
Refs.~\cite{Arvanitaki:2022oby,Huang:2024tog,Gruzinov:2024ciz,Kalia:2024xeq} compute the density modulation of the neutrino background near Earth's surface at the level of the nonrelativistic field equations with a spatially varying index of refraction, which should again be equivalent to the formalism presented here in the respective applicable regimes. App.~\ref{app:nonperturbative} constitutes partial numerical confirmation of both the perturbative results of the main text and of the conclusions of Refs.~\cite{Gruzinov:2024ciz,Kalia:2024xeq}.
While this work was being completed, Ref.~\cite{Blas:2022ovz} came to my attention; their wavepacket approach for neutrinos more closely resembles the kinetic-theory formalism of \Sec{sec:quantum}.

The primary novelty of this work is its more illuminating treatment of wake forces as perturbative classical wave phenomena in \Sec{sec:classical} and within the framework quantized kinetic theory in \Sec{sec:quantum}. The methods presented here should agree with those of the above-mentioned works in their regimes of validity. However, the formalism of this work allows for the efficient computation of the spatial profile of wake forces (\Sec{sec:range}) for phenomenologically realistic background distributions (\Figs{fig:wake_BMB},~\ref{fig:wake_BFD}, and~\ref{fig:form_radial}), including novel oscillation phenomena. \Apps{app:scalars} and~\ref{app:fermions} show that wake forces arise and can be computed for $\emph{any}$ quadratically-coupled field. 
\App{app:nonperturbative} describes the implementation and results of nonperturbative numerical simulations of wake forces, which validate the perturbative methods in this work in $2+1$ and $3+1$ dimensions.

I have shown in \Sec{sec:dmsearches} that wake forces can form the basis for a new class of DM searches with precision-frontier force experiments, in a mass range that is challenging to probe with other methods. Force experiments can thus bridge the gap between traditional DM scattering experiments on the one hand, and DM searches based on coherent interactions to first order in the coupling on the other hand. The optimization of sensors and setups for wake forces should be explored further, and the parameter space of relevant DM models charted out more extensively. Spin-dependent DM wake force searches are an obvious next step. A formal proof of the convergence of the perturbative series, including corrections to the wake potential of $\mathcal{O}(G^3)$, is left to future work. The exquisite accuracy of the leading-order wake potential relative to the nonperturbative simulations in \App{app:nonperturbative} strongly suggests that the expansion parameter is no larger than $\epsilon \sim G m n_\psi / k_0^2$, so that the perturbative series converges whenever the no-screening condition of \Eq{eq:no_screening} is satisfied. I speculate that the agreement with the leading-order wake potential prediction, despite potentially large phase accumulation across large sources, is due a combination of its phase insensitivity and its short-ranged nature in two or more spatial dimensions.

It would be amusing to obtain a positive measurement of wake forces mediated by known elementary particles. Unfortunately, there are not too many suitable light bosons to go around in the SM. The simplest possibility may be force measurements between (electrically neutral) dielectric source and target samples, to which the photon couples quadratically. Wake forces and potentials from neutrinos appear to be prohibitively tiny, no matter the neutrino source (\Sec{sec:neutrino}). An ensemble of ultracold neutrons or atoms/molecules in a cold beam may be more promising fermionic mediators of wake forces. Finally, the phenomena calculated in this work for elementary particles undoubtedly carry over to quasiparticles in condensed matter systems---indeed, they may form the basis for certain types of exotic superconductors~\cite{Kohn:1965zz}, but it would be interesting to study phonons, magnons, and other collective excitations as mediators of wake forces.

\begin{acknowledgments}
    I am grateful for collaboration with Sergei Dubovsky in the initial stages of this project in 2018--2019; he contributed conceptual understanding to \Sec{sec:theory} and model-building aspects of \Sec{sec:dmsearches}. Riccardo Rattazzi's questions about the treatment in \Sec{sec:quantum} encouraged me to come up with the simpler description in \Sec{sec:classical}. 
    I enjoyed discussing this work and Ref.~\cite{Arvanitaki:2022oby} with Asimina Arvanitaki and Savas Dimopoulos, even though none of us fully appreciated the extent of the overlap of our work until recently. They raised concerns about the possible breakdown of the perturbative expansion, and pushed me to produce \App{app:nonperturbative}.
    I thank Mitrajyoti Ghosh, Yuval Grossman, Walter Tangarife, Xun-Jie Xu, and Bingrong Yu for comments, encouragement, and discussions of Ref.~\cite{Ghosh:2022nzo}.
    Hannah Day, Da Liu, Markus Luty, and Yue Zhao gracefully coordinated a companion paper~\cite{Day:2023mkb} to this work that expounds on the elastic scattering and screening effects from \Secs{sec:elastic} \&~\ref{sec:screening} in much greater detail; they had also started considering the concepts of wake forces. 
    Credit to Olivier Simon for pointing out the connections to Ref.~\cite{Montgomery:1968}.
    I also benefited from discussions with Masha Baryakhtar, Kyle Cranmer, David Dunsky, Marco Farina, Simon Foreman, Marat Freytsis, Marios Galanis, Isabel Garcia Garcia, Andrei Gruzinov, Matthew Kleban, Simon Knapen, Marius Kongs{\o}re, Paul Langacker, Matthew Low, Duccio Pappadopulo, Olivier Simon, Philip S{\o}rensen, Neal Weiner, Zachary Weiner, Tien-Tien Yu, and especially Junwu Huang, who joined me on (failed) attempts to make the wake force from high-energy monochromatic neutrinos observable.
    
    This work is supported by the National Science Foundation under Grant PHY-2210551. 
    The Center for Computational Astrophysics at the Flatiron Institute is supported by the Simons Foundation. 
    This work was performed in part at the Institute for Nuclear Theory at the University of Washington, supported in part by the U.S. Department of Energy grant No.~DE-FG02-00ER41132; and the Aspen Center for Physics, supported by National Science Foundation grants PHY-1607611 and PHY-2210452.
\end{acknowledgments}

\appendix

\section{Other Scalar Interactions} \label{app:scalars}

In this appendix, I present the derivation of the wake force for two interactions other than that of $\phi^2 \bar{\psi} \psi$ in \Eq{eq:L_scalar}. The form of the monopole-monopole wake force by a \emph{complex} scalar $\Phi$ is identical to that of a real scalar $\phi$. This extension is (almost) trivial since a complex scalar can be decomposed into two real scalars, but it shows that the wake force does not depend on particle-antiparticle asymmetry. The second example adds a pseudoscalar interaction to the Lagrangian of \Eq{eq:L_scalar}, which leads to an additional monopole-dipole and a dipole-dipole wake force.

\subsection{Complex scalar} \label{app:complex_scalar}
For simplicity, I follow the steps of the classical derivation in \Sec{sec:classical} (the quantized treatment gives identical results) and consider a complex scalar field $\Phi$ with Lagrangian
\begin{align}
\mathcal{L} = (\partial \Phi)(\partial \Phi^\dagger) - m^2\Phi\Phi^\dagger - G m \Phi \Phi^\dagger \bar{\psi} \psi, \label{eq:L_complex_scalar}
\end{align}
and corresponding equation of motion for $\Phi$, with a source particle at the origin:
\begin{align}
(\Box + m^2)\Phi = - G m \Phi \deltathree(\vect{x}). \label{eq:Phi_eom}
\end{align}
As before, decompose the field into a $G$-independent background and a spherical perturbation linear in $G$ from a source particle at the origin:
$\Phi = \Phi_0 e^{-i(\omega t-\vect{k}_0\cdot \vect{x})} + \delta \Phi(t,\vect{x})$ with $\omega^2 = m^2 + \vect{k}_0^2$. The solution to the equation of motion to leading (first) order in $G$ is:
\begin{align}
\delta \Phi(t,\vect{x}) &= G m \Phi_0 \int  \ddbar^4p \, e^{-i p \cdot x} \frac{\deltabar(p_0-\omega)}{(p_0+i\varepsilon)^2-\vect{p}^2 - m^2 }\nonumber\\
&= G m\Phi_0 e^{-i \omega t} \int  \ddbar^3p \, e^{i \vect{p}\cdot \vect{x}} \frac{1}{-\vect{p^2}+\vect{k}_0^2 + i\varepsilon}\nonumber\\
&= G m \Phi_0 \frac{-i}{r}  \int \frac{\dd p}{(2\pi)^2} p \frac{e^{i pr }e^{-i \omega t}}{-p^2+\vect{k}_0^2 + i\varepsilon}\nonumber\\
&=  -\frac{G m}{4\pi r} \Phi_0 e^{-i(\omega t-kr)}, \label{eq:deltaPhi_complex}
\end{align}
where I show more intermediate steps than for the real-scalar derivation in \Sec{sec:classical}.
The wake potential is the obtained by retaining the term of the potential energy $V = G m \Phi \Phi^\dagger$ of $\psi$ to second order in $G$, which after time averaging is:
\begin{align}
\langle V(t,\vect{x}) \rangle_t = -\frac{G^2}{4\pi r} 2 m^2 \Phi_0 \Phi_0^\dagger \cos(|\vect{k}_0|r - \vect{k}_0\cdot \vect{x}). \label{eq:V_complex_1}
\end{align}
This has the same form as \Eq{eq:V_1} for the real scalar, noting that the energy density in the nonrelativistic limit is $\rho = 2 m^2 \Phi_0 \Phi_0^\dagger$. The quantized treatment of the complex scalar yields consistent results, and is identical to \Eqs{eq:V_wake_scalar} \&~\ref{eq:form_general} with the replacement $n \to (n + \bar{n})$---the wake forces of particles and antiparticles are additive. 

The results above are a trivial extension from the case of a real scalar, since a complex scalar can be decomposed into two real scalars, e.g.~$\Phi = (\phi_1 + i \phi_2)/\sqrt{2}$ so $G m \Phi^\dagger \Phi \bar{\psi} \psi = (G m / 2) (\phi_1^2 + \phi_2^2) \bar{\psi} \psi$, each of which produce an identical wake force. It is nevertheless instructive that wake forces do not depend on particle-antiparticle asymmetry, a result that also holds for fermion-mediated wake forces as shown in \App{app:fermions}, in contrast to fermion potentials linear in $G$ (e.g.~\Eq{eq:Omega_CNB_3}).

\subsection{Pseudoscalar interactions} \label{app:pseudoscalar}
Consider the quadratic interactions of a real scalar or pseudoscalar $\phi$ with both the scalar and pseudoscalar currents of $\psi$
\begin{align}
\mathcal{L} = \frac{1}{2}(\partial \phi)^2 - \frac{1}{2}m^2\phi^2 - \frac{m \phi^2}{2} \left(G_s \bar{\psi} \psi + G_p \bar{\psi} i\gamma^5 \psi
    \right), \label{eq:L_pseudoscalar}
\end{align}
with couplings $G_s$ and $G_p$, respectively. In addition to the monopole-monopole wake force proportional to $G_s^2$ derived in \Sec{sec:theory}, the addition of the pseudoscalar interaction also generates a monopole-dipole wake force proportional to $G_s G_p$ and a dipole-dipole one proportional to $G_p^2$, which will be derived in the quantized treatment below.

With the inclusion of the pseudoscalar interaction, the amplitude corresponding to the diagrams in \Fig{fig:feyn_scalar} is:
\begin{alignat}{4} 
    &i \mathcal{M} = & & \\
    &-i m^2 \int \ddbar^3 k \, \frac{n f(\vect{k})}{2 E_k}
    \hspace{-0.8em} &       &\left[\bar{u}_1^{s_1'}(p_1 - q) (G_s + G_p i\gamma^5) u_1^{s_1}(p_1)\right] \nonumber \\
    & &\times &\left[\frac{1}{(q+k)^2-m^2}+ \frac{1}{(q-k)^2 - m^2} \right] \nonumber \\
    & &\times &\left[\bar{u}_2^{s_2'}(p_2) (G_s + G_p i\gamma^5) u_2^{s_2}(p_2 + q) \right]. \nonumber
\end{alignat}
Treating the $\psi$ fermions nonrelativistically, the identities of \Eqs{eq:id_spinor_5} \&~\ref{eq:id_spinor_6} can be used to simplify the amplitude:
\begin{alignat}{4}
    &\frac{i \mathcal{M}}{(2 m_{\psi_1})(2 m_{\psi_2})} \simeq  & & \\
    &- i m^2 \int \ddbar^3 k \, \frac{n f(\vect{k})}{2 E_k} 
        & & \left[G_s + G_p \frac{i \vect{q} \cdot \vect{\sigma}_1}{2 m_{\psi_1}}\right]
            \left[G_s - G_p \frac{i \vect{q} \cdot \vect{\sigma}_2}{2 m_{\psi_2}}\right] \nonumber \\
    &   &\hspace{-3em} \times & \left[\frac{1}{(q+k)^2-m^2}+ \frac{1}{(q-k)^2 - m^2} \right]. \nonumber 
\end{alignat}
Setting the energy transfer between the fermions to zero ($q^\mu = (q^0,\vect{q}) \simeq 0$), selecting the $i \varepsilon$ prescription for the retarded propagator, and using the external four-momentum to cancel the pole ($k^2 = E_k^2 - \vect{k}^2 = m^2$), the propagator denominators are of the form: $(q \pm k)^2 - m^2 \simeq -(\vect{q}^2 \pm 2 \vect{q} \cdot \vect{k} \mp i \varepsilon)$. The Fourier transform of the potential is related to the amplitude as $\widetilde{V}(\vect{q}) = -  \mathcal{M}/[(2m_{\psi_1})(2m_{\psi_2})]$, so the potential in coordinate space equals:
\begin{alignat}{2}
    V(\vect{x}) &= - m^2 \int \ddbar^3 k \, \frac{n f(\vect{k})}{2 E_k}  \int \ddbar^3 q  e^{i \vect{q}\cdot\vect{x}} \label{eq:V_ss_sp_pp}\\
    &\phantom{\simeq} \times \frac{
    \left[G_s + G_p \frac{i \vect{q} \cdot \vect{\sigma}_1}{2 m_{\psi_1}}\right]
    \left[G_s - G_p \frac{i \vect{q} \cdot \vect{\sigma}_2}{2 m_{\psi_2}}\right]
    }
    {\vect{q}^2 + 2 \vect{q} \cdot \vect{k} - i \varepsilon} 
    + \cc \nonumber
\end{alignat}
The $G_s^2$ piece of this expression, call it $V_{ss}$, matches \Eq{eq:V_3}.

\paragraph*{Monopole-dipole wake potential---}
Let $\psi_1$ be the (unpolarized) source and $\psi_2$ the (polarized) target. The $G_s G_p$ piece of the potential can be rewritten as:
\begin{alignat}{2}
    V(\vect{x}) &\supset - \frac{G_s G_p n m}{4 m_\psi} \int \ddbar^3 k \, \frac{f(\vect{k})}{E_k}  e^{-i \vect{k} \cdot \vect{x}} \\
    &\phantom{\supset} \int \ddbar^3 q  e^{i \vect{q} \cdot\vect{x}} (-i) \frac{(\vect{q}-\vect{k}) \cdot \vect{\sigma}_2}{\vect{q}^2 -\vect{k}^2 - i \varepsilon} + \cc, \nonumber
\end{alignat}
after shifting the integration variable $\vect{q} \to \vect{q} - \vect{k}$ in the first equality. Then using the integral formulae of \Eqs{eq:integral_1} \&~\ref{eq:integral_2}, the monopole-dipole potential is:
\begin{alignat}{2}
    &V_{sp}(\vect{x},\vect{\sigma}_2) = - \frac{G_s G_p m n}{4\pi r} \frac{\vect{\sigma}_2}{2m_{\psi_2}} \cdot \int \ddbar^3 k \, \frac{f(\vect{k})}{E_k}  \label{eq:V_wake_sp}
    \\
    &\hspace{1.2em}\times \left\lbrace \frac{\hat{\vect{x}}}{r} \cos(|\vect{k}|r - \vect{k}\cdot\vect{x}) + \left[|\vect{k}| \hat{\vect{x}} - \vect{k} \right] \sin(|\vect{k}|r - \vect{k}\cdot \vect{x})  \right\rbrace \nonumber.
\end{alignat}
As expected, the monopole-dipole potential is related to the monopole-monopole potential as:
\begin{alignat}{2}
    V_{sp}(\vect{x},\vect{\sigma}_2) = \frac{G_p}{G_s} \frac{-\vect{\sigma}_2 \cdot \vect{\nabla}{}}{2 m_{\psi_2}} V_{ss}(\vect{x}), \label{eq:V_sp_ss_relation}
\end{alignat}
which can also be seen directly from \Eq{eq:V_ss_sp_pp}.
This relation implies that one can compute the monopole-monopole wake potential first (including the intensive integrals over $\vect{k}$ to obtain the form factors), and then obtain the monopole-dipole wake potential simply by taking the gradient of the monopole-monopole one. The monopole-dipole potential thus has a universal scaling as $1/r^2$ at short distances, and falls off with one \emph{extra} power of $1/r$ compared to the monopole-monopole potential at large distances, e.g.~$V_{sp} \propto 1/r^4$ for BMB and BFD distributions, where it is effectively short ranged.

\paragraph*{Dipole-dipole wake potential---}
Likewise, the dipole-dipole wake potential for $\psi_2$ as sourced by $\psi_1$ can be written as:
\begin{alignat}{2}
    V_{pp}(\vect{x},\vect{\sigma}_1,\vect{\sigma}_2) = \left(\frac{G_p}{G_s}\right)^2 \frac{+\vect{\sigma}_1 \cdot \vect{\nabla}{}}{2 m_{\psi_1}}\frac{-\vect{\sigma}_2 \cdot \vect{\nabla}{}}{2 m_{\psi_2}} V_{ss}(\vect{x}), \label{eq:V_pp_ss_relation}
\end{alignat}
and exhibits a universal $1/r^3$ short-distance scaling.

\section{Fermions} \label{app:fermions}

Wake forces can be mediated by spin-$1/2$ particles as well as spin-0 particles. In this appendix, I enumerate some of the leading interactions of Dirac fermions with SM fermion currents of scalar ($\bar{\psi} \psi$), vector ($\bar{\psi} \gamma^\mu \psi$), and axial-vector ($\bar{\psi} \gamma^\mu \gamma^5 \psi$) type, and calculate their resulting wake potentials. I comment on the extension to Majorana fermions at the end of this appendix.

\subsection{Dirac fermion, scalar current} \label{app:Dirac_ss}
The simplest example of a fermion-mediated wake force is that of a Dirac fermion $\chi$ with a scalar current $\bar{\psi} \psi$:
\begin{align}
\mathcal{L} = \bar{\chi} (i\slashed{\partial}-m)\chi - \frac{G_s}{2} \bar{\chi} \chi \bar{\psi} \psi. \label{eq:L_Dirac_ss}
\end{align}
Like for the scalar field in \Eq{eq:phi_quant}, quantize the Dirac field as:
\begin{align}
\chi(x) = \int \frac{\ddbar^3 k}{\sqrt{2E_k}}\, \sum_s \left[a_{\vect{k}}^s u^s(k) e^{-ik\cdot x}+ b_{\vect{k}}^{s\dagger} v^s(k) e^{ik\cdot x}\right] \label{eq:chi_quant}
\end{align}
with annihilation (creation) operators $a_{\vect{k}}$ ($a_{\vect{k}}^\dagger$) for particles and $b_{\vect{k}}$ ($b_{\vect{k}}^\dagger$) for particles and antiparticles, respectively. The corresponding spinor solutions $u^s(k)$ and $v^s(k)$ satisfy the standard identities of \Eqs{eq:id_spinor_1} \&~\ref{eq:id_spinor_2}. The background is taken to be a mixed state with a number density $n$ of $\chi$ particles and a number density $\bar{n}$ of $\bar{\chi}$ antiparticles with unpolarized spins and momentum distribution $f(\vect{k})$:
\begin{align}
\left\langle a_{\vect{k}'}^{s'\dagger} a_{\vect{k}}^s \right\rangle &= n f(\vect{k}) \deltabarthree(\vect{k}'-\vect{k})\frac{\delta^{s's}}{2}; \label{eq:a_a_dag_Dir}\\
\left\langle b_{\vect{k}'}^{s'\dagger} b_{\vect{k}}^s \right\rangle &= \bar{n} f(\vect{k}) \deltabarthree(\vect{k}'-\vect{k})\frac{\delta^{s's}}{2}. \label{eq:b_b_dag_Dir}
\end{align}
The total energy density in the medium $\chi$ and $\bar{\chi}$ particles is then just: 
\begin{align}
    \hspace{-0.4em}\rho_\chi = \left \langle  \bar{\chi} \left(-i \vect{\gamma} \cdot \vect{\nabla} + m \right) \chi \right \rangle = \int \ddbar^3k \, f(\vect{k}) E_k (n+\bar{n}).
\end{align}

\begin{figure}[t]
    \begin{tikzpicture}
        \begin{feynman}
          \vertex (i1) {\(\psi_1\)};
          \vertex [above=of i1] (v1);
          \vertex [above=of v1] (f1) {\(\psi_1\)};
          \vertex [right=of i1] (i2) {\(\psi_2\)};
          \vertex [above=of i2] (v2);
          \vertex [above=of v2] (f2) {\(\psi_2\)};
          \vertex [left=1cm of i1]  (is) {\(\chi\)};
          \vertex [right=1cm of f2] (fs) {\(\chi\)};
          \diagram* {
          (i1) -- [fermion, very thick] (v1) -- [fermion, very thick] (f1),
          (i2) -- [fermion, very thick] (v2) -- [fermion, very thick] (f2),
          (is) -- [fermion] (v1) -- [fermion, edge label=\(\chi\), momentum'=\(q+k\)] (v2) -- [fermion] (fs)
          };
        \end{feynman}
      \end{tikzpicture} 
      \quad
      \begin{tikzpicture}
        \begin{feynman}
          \vertex (i1) {\(\psi_1\)};
          \vertex [above=of i1] (v1);
          \vertex [above=of v1] (f1) {\(\psi_1\)};
          \vertex [right=of i1] (i2) {\(\psi_2\)};
          \vertex [above=of i2] (v2);
          \vertex [above=of v2] (f2) {\(\psi_2\)};
          \vertex [right=1cm of i2]  (is) {\(\chi\)};
          \vertex [left=1cm of f1] (fs) {\(\chi\)};
          \diagram* {
          (i1) -- [fermion, very thick] (v1) -- [fermion, very thick] (f1),
          (i2) -- [fermion, very thick] (v2) -- [fermion, very thick] (f2),
          (is) -- [fermion] (v2) -- [fermion, edge label'=\(\chi\), reversed momentum=\(q-k\)] (v1) -- [fermion] (fs)
          };
        \end{feynman}
      \end{tikzpicture} 
      
      \begin{tikzpicture}
        \begin{feynman}
          \vertex (i1) {\(\psi_1\)};
          \vertex [above=of i1] (v1);
          \vertex [above=of v1] (f1) {\(\psi_1\)};
          \vertex [right=of i1] (i2) {\(\psi_2\)};
          \vertex [above=of i2] (v2);
          \vertex [above=of v2] (f2) {\(\psi_2\)};
          \vertex [left=1cm of i1]  (is) {\(\bar{\chi}\)};
          \vertex [right=1cm of f2] (fs) {\(\bar{\chi}\)};
          \diagram* {
          (i1) -- [fermion, very thick] (v1) -- [fermion, very thick] (f1),
          (i2) -- [fermion, very thick] (v2) -- [fermion, very thick] (f2),
          (is) -- [anti fermion] (v1) -- [anti fermion, edge label=\(\bar{\chi}\), momentum'=\(q+k\)] (v2) -- [anti fermion] (fs)
          };
        \end{feynman}
      \end{tikzpicture} 
      \quad
      \begin{tikzpicture}
        \begin{feynman}
          \vertex (i1) {\(\psi_1\)};
          \vertex [above=of i1] (v1);
          \vertex [above=of v1] (f1) {\(\psi_1\)};
          \vertex [right=of i1] (i2) {\(\psi_2\)};
          \vertex [above=of i2] (v2);
          \vertex [above=of v2] (f2) {\(\psi_2\)};
          \vertex [right=1cm of i2]  (is) {\(\bar{\chi}\)};
          \vertex [left=1cm of f1] (fs) {\(\bar{\chi}\)};
          \diagram* {
          (i1) -- [fermion, very thick] (v1) -- [fermion, very thick] (f1),
          (i2) -- [fermion, very thick] (v2) -- [fermion, very thick] (f2),
          (is) -- [anti fermion] (v2) -- [anti fermion, edge label'=\(\bar{\chi}\), reversed momentum=\(q-k\)] (v1) -- [anti fermion] (fs)
          };
        \end{feynman}
      \end{tikzpicture} 
\caption{Diagrams contributing to the wake force in a medium of Dirac fermions $\chi$ and antifermions $\bar{\chi}$.} \label{fig:feyn_fermion}
\end{figure}
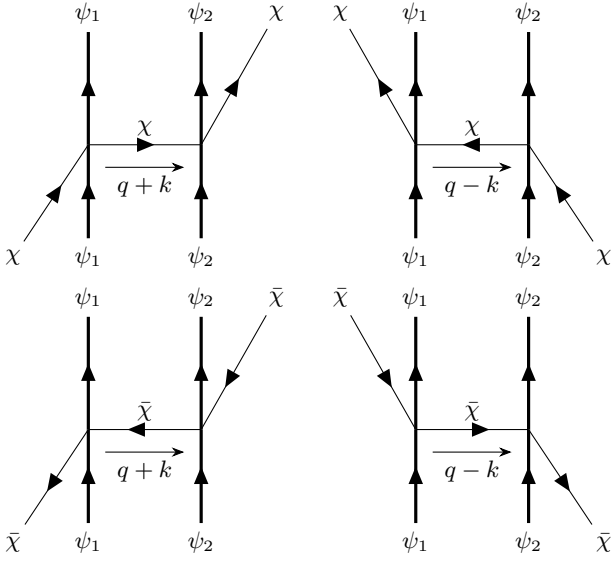

The Feynman diagrams for a fermionic wake force are the same as for the spin-0 equivalent of \Fig{fig:feyn_scalar}, with the addition of a pair of diagrams for the antiparticles $\bar{\chi}$, all shown in \Fig{fig:feyn_fermion}. The combined matrix element evaluates to:
\begin{align}
&\frac{i\mathcal{M}}{(2m_\psi)^2} =- i \frac{G_s^2}{8} \int \ddbar^3k\, \frac{f(\vect{k})}{2E_k} \sum_s \label{eq:mat_fermion_ss}\\
&\hspace{1em} \bigg\lbrace + n \bar{u}^s(k)\left[\frac{\slashed{q}+\slashed{k}+m}{(q+k)^2-m^2}+ \frac{-\slashed{q}+\slashed{k}+m}{(q-k)^2 - m^2} \right]u^s(k) \nonumber \\
&\hspace{1em}\phantom{\bigg\lbrace} - \bar{n} \bar{v}^s(k)\left[\frac{-\slashed{q}-\slashed{k}+m}{(q+k)^2-m^2}+ \frac{\slashed{q}-\slashed{k}+m}{(q-k)^2 - m^2} \right]v^s(k)\bigg\rbrace.\nonumber
\end{align}
The signs require careful attention. The four diagrams of \Fig{fig:feyn_fermion}---top left, top right, bottom left, bottom right---correspond to the contractions with the ``vacuum'' $|\Omega\rangle$, in the same order:
\begin{alignat}{2}
    &\langle
        \wick{
            \c1{\Omega}
            \vert
            \c2{\bar{\chi}}_1 \c3{\chi}_1 \c1{\bar{\chi}}_2 \c2{\chi}_2
            \vert
            \c3{\Omega}
        }
    \rangle \sim +1,  \label{eq:contract_1}\\
    &\langle
    \wick{
        \c1{\Omega}
        \vert
        \c1{\bar{\chi}}_1 \c2{\chi}_1 \c2{\bar{\chi}}_2 \c3{\chi}_2
        \vert
        \c3{\Omega}
    }
    \rangle \sim +1,  \label{eq:contract_2} \\
    &\langle
        \wick{
            \c1{\Omega}
            \vert
            \c3{\bar{\chi}}_1 \c2{\chi}_1 \c2{\bar{\chi}}_2 \c1{\chi}_2
            \vert
            \c3{\Omega}
        }
    \rangle \sim -1,   \label{eq:contract_3} \\
    &\langle
        \wick{
            \c1{\Omega}
            \vert
            \c2{\bar{\chi}}_1 \c1{\chi}_1 \c3{\bar{\chi}}_2 \c2{\chi}_2
            \vert
            \c3{\Omega}
        }
    \rangle \sim -1;  \label{eq:contract_4}
\end{alignat}
where $\bar{\chi}_1 \chi_1$ is the interaction of the current with $\psi_1$, and $\bar{\chi}_2 \chi_2$ with $\psi_2$. Only the signs are evaluated in \Eqs{eq:contract_1}--\ref{eq:contract_4}. The top right diagram of \Fig{fig:feyn_fermion} has no additional minus sign, because no anticommutators are needed to perform the contractions with the vacuum for the incoming and outgoing particles, nor for the internal propagator $ \contraction[0.5ex]{}{\chi}{}{\bar{\chi}} \chi \bar{\chi}$. For the top left diagram and the contractions in \Eq{eq:contract_1}, one needs 2 anticommutators to move $\bar{\chi}_2$ to the left, then 1 to move $\chi_1$ to the right, and finally 1 to swap $\bar{\chi}_1$ and $\chi_2$, for an even number of anticommutators. For the bottom right diagram and \Eq{eq:contract_4}, 1 anticommutator to bring $\chi_1$ leftward, 1 to bring $\bar{\chi}_2$ rightward, and 1 to swap $\bar{\chi}_1$ and $\chi_2$, for an odd number of anticommutators and thus an extra minus sign. The bottom left diagram and \Eq{eq:contract_3} are similarly negative. This explains the extra minus sign for the bottom (antiparticle) diagrams in \Fig{fig:feyn_fermion}, reflected in the last line of \Eq{eq:mat_fermion_ss}.

The the matrix element is similar to that of the scalar case, with the addition of spinor numerators such as $\sum_s \bar{u}^s(k) \left( \slashed{q}+\slashed{k} + m \right) u^s(k) = - 4 \vect{q}\cdot\vect{k} + 8 m^2$ for the first amplitude. The monopole-monopole wake potential is:
\begin{align}
    V(\vect{x}) & = -\frac{G_s^2}{2} \int \ddbar^3k\, \frac{f(\vect{k})}{2E_k} (n+\bar{n}) \\
    & \phantom{= -\frac{G^2}{2}}  \times \int \ddbar^3q\, e^{i\vect{q}\cdot\vect{x}} \frac{-\vect{q}\cdot\vect{k}+2m^2}{\vect{q}^2+2\vect{q}\cdot\vect{k}-i\varepsilon} + \cc \nonumber \\
    &= - \frac{G_s^2 m (n + \bar{n})}{4\pi r} \mathcal{F}(\vect{x}), \label{eq:V_Dirac_ss}
\end{align}
with the form factor appropriate for fermions:
\begin{alignat}{2}
    \mathcal{F}(\vect{x}) &\equiv \int \ddbar^3k\,  \frac{f(\vect{k})}{E_k/m} \label{eq:form_Dirac_ss} \\
    &\phantom{\equiv} \times \Bigg \lbrace 
    \cos(|\vect{k}|r - \vect{k}\cdot\vect{x}) 
    \left[1 +  \frac{|\vect{k}| (|\vect{k}| + \vect{k}\cdot \hat{\vect{x}})}{2m^2} \right] \nonumber \\
    &\phantom{\equiv \times \Bigg \lbrace} \hspace{6em} + 
    \sin(|\vect{k}|r - \vect{k}\cdot\vect{x}) 
    \left[\frac{\vect{k}\cdot \hat{\vect{x}}}{2m^2 r} \right] \Bigg \rbrace. \nonumber
\end{alignat}
In the nonrelativistic limit $|\vect{k}| \ll m$, this form factor is the same as the one for the monopole-monopole wake potential mediated by scalars in \Eq{eq:form_general}. Similarly, the overall fermionic monopole-monopole wake potential of \Eq{eq:V_Dirac_ss} is the same as the scalar one of \Eq{eq:V_wake_scalar} in the coupling normalizations of this work (\Eqs{eq:L_scalar} \&~\ref{eq:L_Dirac_ss}), with fermions and antifermions contributing additively. For nonrelativistic particles at low occupation numbers, there is thus no material difference between monopole-monopole wake forces mediated by fermions and scalars.

\subsection{Four-Fermi interactions of Dirac fermions}  \label{app:four_fermi}
Consider the neutral-current four-Fermi interaction of a single Dirac neutrino in \Eq{eq:L_nu_neutral}, first for a single mass eigenstate. The matrix element for the wake potential diagrams of \Fig{fig:feyn_fermion} is similar to \Eq{eq:mat_fermion_ss}, except the spinor numerators are dressed with additional gamma matrices. Firstly, the linear mass term in the numerator of the fermion propagator does not contribute because the interactions preserve chirality, i.e.~$(1-\gamma^5) \gamma^\nu (1-\gamma^5) = 0$. Secondly, neutrinos and antineutrinos contribute additively as in all previous cases. The full wake potential between two fermions $\psi_1$ and $\psi_2$ with vector and axial couplings $g^\text{V}_1$, $g^\text{A}_1$, $g^\text{V}_2$, $g^\text{A}_2$, and spins $\vect{\sigma}_1$, $\vect{\sigma}_2$ is thus:
\begin{align}
    V(\vect{x},\vect{\sigma}_1,\vect{\sigma}_2) &= -\frac{G_F^2 (n+\bar{n})}{4} \int \frac{\ddbar^3 k}{2E_k} f(\vect{k}) e^{-i\vect{k}\cdot\vect{x}} \\
    &\phantom{=} \times \int \ddbar^3q\, e^{i\vect{q}\cdot\vect{x}} \frac{q_\rho + E_k \delta_\rho^0}{\vect{q}^2 - \vect{k}^2 - i\varepsilon} \mathcal{I}^\rho +\cc, \nonumber
\end{align}
where the $\mathcal{I}$ vector and $\Gamma$ tensors are defined as:
\begin{alignat}{2}
    \mathcal{I}^\rho 
    &\equiv g^\text{V}_1 g^\text{V}_2 \Gamma^{0\rho 0} + g^\text{A}_1 g^\text{A}_2 \sigma^i_{1}\sigma^j_{2} \Gamma^{j\rho i} \nonumber\\
    &\phantom{\equiv}+ g^\text{V}_1 g^\text{A}_2 \sigma^j_{2} \Gamma^{j\rho 0} + g^\text{A}_1 g^\text{V}_2 \sigma^i_{1} \Gamma^{0\rho i} ;\\
    \Gamma^{\nu \rho \kappa} 
    &\equiv \sum_s \bar{u}^s(k) \gamma^\nu(1-\gamma^5)\gamma^\rho \gamma^\kappa (1-\gamma^5) u^s(k) \\
    &= 4 k_\mu \text{tr} \left\lbrace \sigma^\mu \bar{\sigma}^\nu \sigma^\rho \bar{\sigma}^\kappa \right\rbrace \\
    &=8 \left(k^\nu \eta^{\rho \kappa} - k^\rho \eta^{\nu \kappa} + k^\kappa \eta^{\nu \rho} + i k_\mu \epsilon^{\mu \nu \rho \kappa}\right).
\end{alignat}
\paragraph*{Monopole-monopole wake potential---} Using the identity $\Gamma^{0\rho0} = 8 \left(2 E_k \eta^{\rho 0} - k^\rho \right)$, the monopole-monopole wake potential is found to be quite similar to that of the quadratic interaction with the scalar current in \Eq{eq:V_Dirac_ss}:
\begin{align}
    \hspace{-0.2em}
    V(\vect{x}) &= -2G_F^2 g_1^\text{V} g_2^\text{V} (n+\bar{n}) \\
    &\phantom{=} \hspace{-0.2em} \times \int \frac{\ddbar^3k}{2E_k}\, f(\vect{k})  e^{-i\vect{k}\cdot \vect{x}} \hspace{-0.15em} \int \ddbar^3q\, e^{i\vect{q}\cdot\vect{x}} \frac{\vect{q}\cdot\vect{k} + E_k^2}{\vect{q}^2-\vect{k}^2-i\varepsilon} + \cc \nonumber \\
    &= - \frac{2 G_F^2 g_1^\text{V} g_2^\text{V} m(n+\bar{n})}{4\pi r} \mathcal{F}_{VV}(\vect{x}) \label{eq:V_Dirac_VV_2}
\end{align}
with the vector-vector form factor:
\begin{alignat}{2}
    \mathcal{F}_{VV}(\vect{x}) &\equiv \int\frac{\ddbar^3k}{E_k/m}\, f(\vect{k})  \label{eq:form_Dirac_VV}\\
    &\phantom{\equiv}\times \Bigg\lbrace \cos(|\vect{k}| r - \vect{k}\cdot \vect{x} )\left[1 + \frac{|\vect{k}| (|\vect{k}| - \vect{k}\cdot \hat{\vect{x}})}{m^2}\right] \nonumber \\
    &\phantom{\equiv \times \Bigg\lbrace} \hspace{6em} - \sin(|\vect{k}| r - \vect{k}\cdot \vect{x} ) \left[ \frac{\vect{k} \cdot \hat{\vect{x}}}{m^2 r} \right] \Bigg\rbrace. \nonumber
\end{alignat}
In the nonrelativistic limit, this form factor reduces to that of \Eqs{eq:form_general} \&~\ref{eq:form_Dirac_ss}.

\paragraph*{Monopole-dipole wake potential---}
Without loss of generality, consider the part of the wake monopole-dipole potential proportional to $g_1^\text{V} g_2^\text{A}$. Use $\Gamma^{j\rho 0} = 8 (k^j \eta^{\rho 0} + E_k \eta^{j \rho} + i k_\mu \epsilon^{\mu j \rho 0})$ to find:
\begin{align}
    V(\vect{x},\vect{\sigma}_2) 
    &= \frac{2 G_F^2 g_1^\text{V} g_2^\text{A} m (n+\bar{n})}{4\pi r} \vect{\sigma}_2 \cdot \vect{\mathcal{F}}_{VA}(\vect{x}), \label{eq:V_Dirac_VA}
\end{align}
where the form factor is now a vector function:
\begin{align}
    &\vect{\mathcal{F}}_{VA}(\vect{x}) =  \int\frac{\ddbar^3k}{E_k/m}\, f(\vect{k}) \label{eq:form_Dirac_VA}\\
    &\times \Bigg\lbrace\cos(|\vect{k}| r - \vect{k}\cdot \vect{x} )\left[\frac{E_k(\vect{k} + |\vect{k}| \hat{\vect{x}})}{m^2} - \frac{\hat{\vect{x}} \cdot (\vect{k}\times \vect{\sigma}_2)}{m^2 r}\right] \nonumber \\
    &\phantom{\times \Bigg\lbrace} \hspace{2em} - \sin(|\vect{k}| r - \vect{k}\cdot \vect{x} )\left[ \frac{E_k \hat{\vect{x}}}{m^2 r} + \frac{|\vect{k}| (\hat{\vect{x}} \times \vect{k})}{m^2} \right] \Bigg\rbrace. \nonumber
\end{align}
In the nonrelativistic limit and $r \lesssim 1/|\vect{k}_0|$, this form factor generically has a magnitude of order the typical velocity $|\vect{k}_0|/m$ of the medium.

\subsection{Majorana fermions}
Suppose the fermions $\chi$ are their own antiparticles, and quantize the theory with creation and annihilation operators $a_{\vect{k}}^\dagger$ and $a_{\vect{k}}$:
\begin{align}
    \chi(x) = \int \frac{\ddbar^3 k}{\sqrt{2E_k}}\, \sum_s
    \left[ a_{\vect{k}}^s u^s(k) e^{-ik\cdot x}+ a_{\vect{k}}^{s\dagger} i \gamma_2 u^s(k) e^{ik\cdot x} \right].
\end{align}
Assume the modes are populated with number density $n$ and momentum distribution $f(\vect{k})$ according to \Eq{eq:a_a_dag_Dir}.
More contractions are possible for  Majorana fields, namely $ \contraction[0.5ex]{}{\chi}{}{\chi} \chi \chi$ and $\contraction[0.5ex]{}{\bar{\chi}}{}{\bar{\chi}} \bar{\chi} \bar{\chi}$ in addition to the usual  $\contraction[0.5ex]{}{\chi}{}{\bar{\chi}} \chi \bar{\chi}$. 
For quadratic interactions of the form $\bar{\chi} \Gamma \chi$ with $\Gamma = \lbrace 1, i\gamma_5, \gamma_\mu \gamma_5 \rbrace$, these additional contractions can all be ``absorbed'' by conventional factors of $1/2$ in the vertices and propagators. For example, consider the Lagrangian of a Majorana fermion interacting with a scalar current:
\begin{align}
\mathcal{L} = \frac{1}{2}\bar{\chi} (i\slashed{\partial}-m)\chi - \frac{G}{4} \bar{\chi} \chi \bar{\psi} \psi .\label{eq:L_Maj_ss}
\end{align}
This is the analog of \Eq{eq:L_Dirac_ss}, and will yield the \emph{identical} wake potential to \Eq{eq:V_Dirac_ss} with this normalization of $G$ and with the replacement $(n+\bar{n}) \to n$. Vertex structures with $\Gamma = \lbrace  \gamma_\mu, \sigma_{\mu\nu}\rbrace$ yield zero.

\section{Nonperturbative wake} \label{app:nonperturbative}
The primary approximation made in the main text is the $G \to 0$ limit. As argued in \Sec{sec:screening}, this perturbative expansion should converge for ${|G| m n_\psi}/k_0^2 \ll 1$. (The other approximation is the classical $\hbar \to 0$ limit, but \Sec{sec:double_exchange} showed that finite-$\hbar$ effects are negligible at $r \gtrsim \sigma_k^{-1}$ in practice.)
In this appendix, the perturbative method of this work is validated against numerical simulations of the wake induced by a perturber of arbitrary strength and size.

Numerical experiments become more feasible in the nonrelativistic limit, where the ``Compton oscillations'' of the field are integrated out. Specifically, \Eq{eq:EOM_scalar} reduces to a Schr{\" o}dinger equation:
\begin{alignat}{2}
    i \partial_t \Psi(t,\vect{x}) = \frac{-\vect{\nabla}^2+ G m n_\psi(\vect{x})}{2m} \Psi(t,\vect{x}), \label{eq:schrodinger}
\end{alignat}
after substituting $\phi = e^{-i m t} \Psi/ \sqrt{2 m} + \cc$ and ignoring rapid relative oscillations at frequencies of $\mathcal{O}(m)$. The number density of nonrelativistic $\phi$ particles is thus quantified by $n \simeq |\Psi|^2$, and the full nonperturbative potential for $\psi$ particles is:
\begin{alignat}{2}
    V = \frac{G |\Psi|^2}{2}. \label{eq:V_NPT}
\end{alignat}
For illustrative purposes and numerical efficiency, \Eq{eq:schrodinger} is simulated both in $2+1$ and $3+1$ dimensions in what follows.

The simulations are performed in dimensionless time coordinates with $m=1$ and on a unit $n$-torus: $x,y \in [-1/2,1/2)$ for $n=2$, and $x,y,z \in [-1/2,1/2)$ for $n=3$, with identified sides. (The only parametrization-invariant quantities are $G m n_\psi / \vect{k}^2$ and the fractional gradients of $n_\psi$ relative to $|\vect{k}|$; no meaning should be attached to the absolute parameter values except for these ratios.) The initial condition for $\Psi(t=0,\vect{x})$ is that of a Gaussian random field: the Fourier transform of the wavefunction $\widetilde{\Psi}({\vect{k}}) = \int \dd^n x \, e^{-i \vect{k} \cdot \vect{x}} \Psi(\vect{x})$ has an isotropic Maxwell-Boltzmann power spectrum $\langle |\widetilde{\Psi}(\vect{k})|^2 \rangle \propto e^{-\vect{k}^2/2\sigma_k^2}$ and a random phase for each mode $\vect{k}$. 
An $n$-sphere of uniform density $n_\psi$ and radius $R$ is centered at $\vect{x}=0$. 
Space is discretized in steps $\dd x = \dd y = 1/2{,}048$ in $2+1$D and in steps $\dd x = \dd y = \dd z = 1/256$ in $3+1$D. The simulations evolve the initial wavefunction forward in time over the interval $t \in [0,0.1]$ in increments of $\dd t = 5 \times 10^{-6}$ in $2+1$D and $\dd t = 10^{-4}$ in $3+1$D; the density is recorded in each time step only after $t > 0.025$ to avoid initial transients. Forward evolution of the wavefunction is done via a spectral method: each time step, $\Psi(\vect{x})$ is Fourier transformed, evolved as $\widetilde{\Psi}(\vect{k}) \to  e^{-i \vect{k}^2 \mathrm{d}t / 2} \widetilde{\Psi}(\vect{k})$, and then inverse Fourier transformed. Similarly, the phase evolution from the potential $G n_\psi(\vect{x})/2$ is computed each time step, in real space.
The mean and (statistical) standard deviation of the density $n(\vect{x}) = |\Psi(\vect{x})|^2$ are computed over 5{,}000 runs with random initializations of the field. 

\paragraph*{2+1 dimensions---}

\begin{figure}[htb!]
    \centering
    \includegraphics[width=0.45\textwidth]{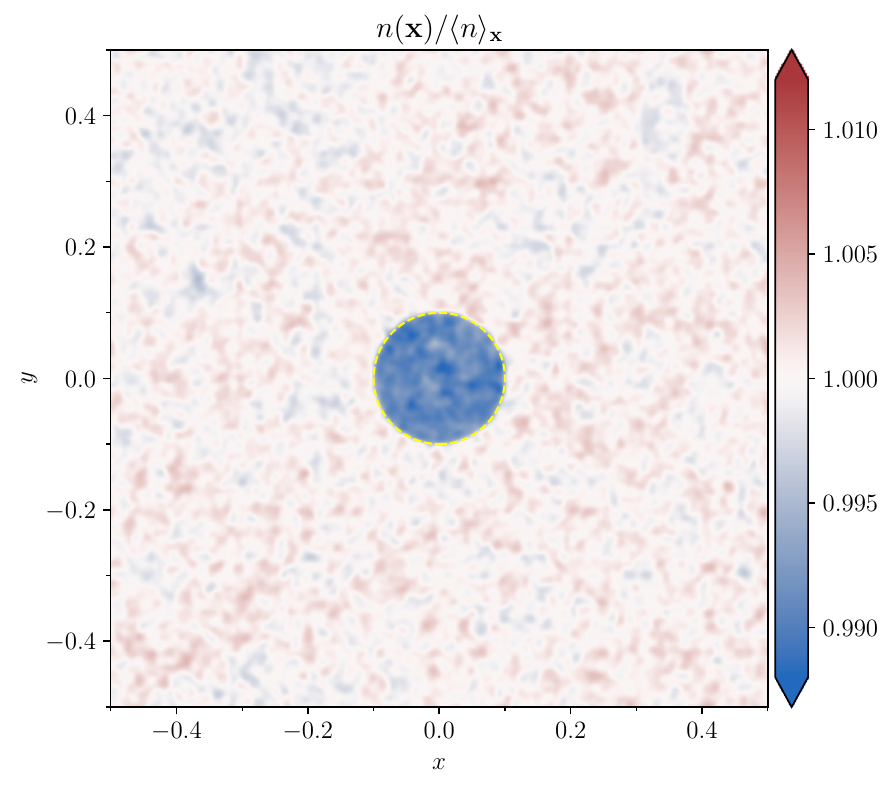}
    \caption{Time-averaged number density $n(\vect{x}) = |\Psi(\vect{x})|^2$ over a 2-torus in $2+1$D simulations of nonrelativistic $\phi$ particles with isotropic Maxwell-Boltzmann distribution parametrized by a momentum spread  $\sigma_k = 100$. The particles are scattered by a uniform disk of radius $R = 0.1$ (dashed yellow outline) with an effective coupling strength $G m n_\psi = +200$ in its interior. The density profile is normalized to its 2-torus average $\langle n \rangle_{\vect{x}}$. Upon taking the mean over many coherence times and simulations, the fractional density profile has permille-level \emph{statistical} density variations, revealing the fractional density modulation of order $-G m n_\psi / 2 \sigma_k^2 = -0.01$ due to the wake of the disk. The coherence length of the density variation/modulation is of $\mathcal{O}(\sigma_k^{-1}$).}
    \label{fig:wake_sim_2D}
\end{figure}

\begin{figure*}[htb!]
    \centering
    \includegraphics[width=0.45\textwidth]{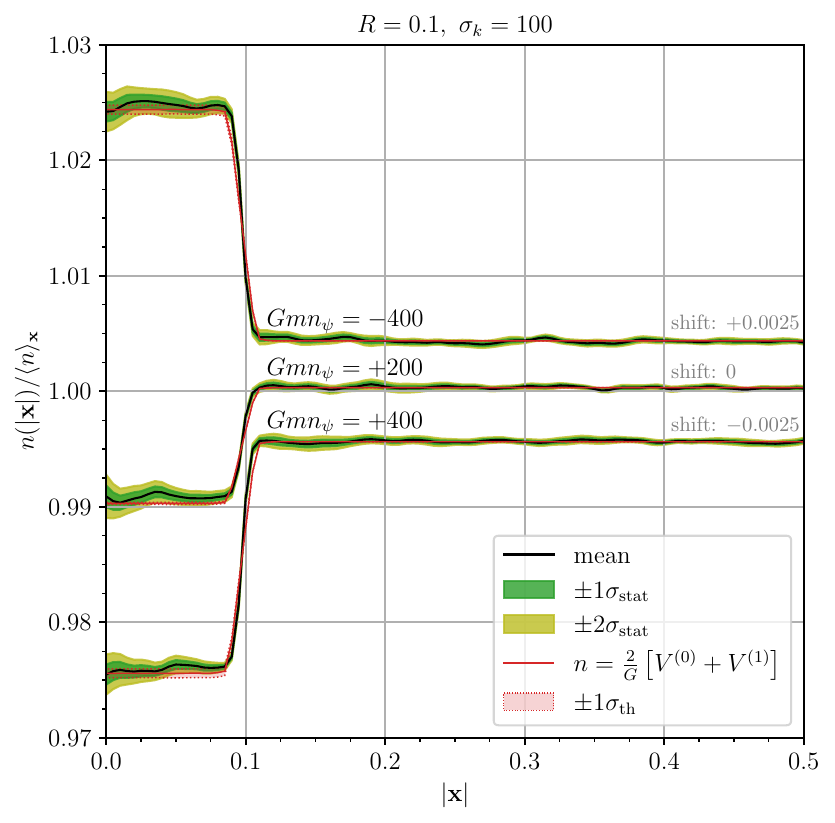}
    \includegraphics[width=0.45\textwidth]{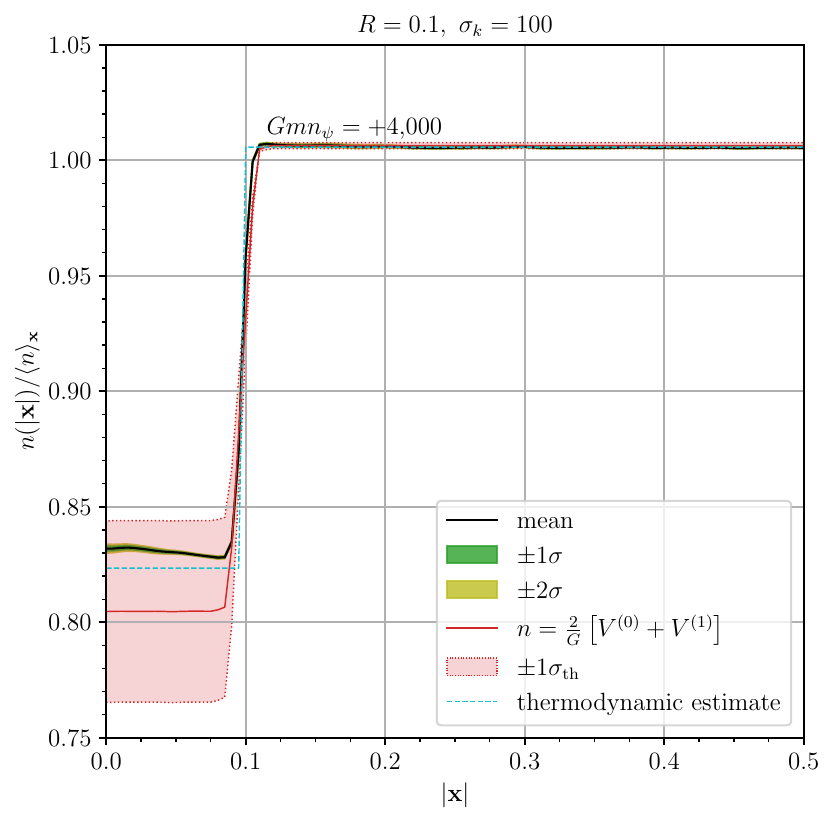}
    \vspace{-1em}
    \caption{Number density $n(|\vect{x}|)$ relative to the 2-torus average $\langle n \rangle_{\vect{x}}$ in $2+1$D as in \Fig{fig:wake_sim_2D} but azimuthally averaged at a distance $|\vect{x}|$ from the center of the disk-shaped barrier/well with $R = 0.1$, and for a MB distribution with $\sigma_k = 100$. The effective coupling inside the disk is varied over $G m n_\psi = \lbrace -400, +200, +400, +4{,}000 \rbrace$, with the first three displayed with vertical shifts of $0.0025$ in the left panel, and the last shown separately in the right panel. Black curves show the time-averaged means over approximately 5{,}000 simulations, and the green (yellow) bands indicate $1\sigma_\mathrm{stat}$ ($2\sigma_\mathrm{stat}$) statistical errors. Red curves depict the leading-order wake potential prediction; red bands include a $1\sigma_\mathrm{th}$ estimated theory error of $\sigma_\mathrm{th} \simeq \epsilon^2 \equiv (G m n_\psi / 2 \sigma_k^2)^2$. The thermodynamic estimate from \Eq{eq:thermo} is included in dashed light blue. The systematic error from partial field initialization inside the barrier is estimated to be $\sigma_\mathrm{sys} \sim \epsilon \pi R^2 \sim 0.006$ for the right panel (and negligible in the left panel) but is not shown.}
    \label{fig:wake_sim_1D}
\end{figure*}

The average density profile $n(\vect{x})$ of the $2+1$D simulations is shown as a function of $\vect{x} = (x,y)$ and $|\vect{x}|$ (azimuthal average) in \Figs{fig:wake_sim_2D} and~\ref{fig:wake_sim_1D}  respectively, and is normalized relative to the average number density $\langle n \rangle_{\vect{x}}$ over the 2-torus. The simulation parameters are $\sigma_k = 100$ and $G m n_\psi(\vect{x}) = \lbrace -400,+200,+400,+4{,}000 \rbrace$ for $|\vect{x}| < R = 0.1$, and zero otherwise. 
The Gaussian random field initialization does not strictly correspond to a pure Maxwell-Boltzmann distribution in the vacuum \emph{outside} the barrier, because a fraction $\pi R^2$ of the total density is initialized \emph{inside} the barrier at a slightly higher energy. I conservatively estimate the systematic error of this effect on $n(x)/\langle n \rangle_x$ to be $\sigma_\mathrm{sys} \sim \epsilon \pi R^2$ inside the barrier, and $\sigma_\mathrm{sys} \sim \epsilon (\pi R^2)^2$ outside. These systematic deviations decrease with smaller fractional volumes occupied by the barrier.

\Eq{eq:schrodinger} can also be solved perturbatively, using the same methods as in the main text. Decompose the field into the perturbative series $\Psi = \Psi^{(0)} + \Psi^{(1)} + \dots$ and similarly for the potential of \Eq{eq:V_NPT}:
\begin{alignat}{2}
    V \simeq V^{(0)} + V^{(1)} = G \frac{|\Psi^{(0)}|^2}{2} + G \, \mathrm{Re} \,\lbrace \Psi^{(0)*} \Psi^{(1)} \rbrace
\end{alignat}
to second order in $G$.
The background wave can always be written as a sum of plane waves $\Psi^{(0)} = \sum_{\vect{\alpha}} \Psi_{\vect{\alpha}} e^{-i (\omega_\alpha t - \vect{k}_\alpha \cdot \vect{x})}$, i.e.~the zeroth-order solutions of the Schr{\" o}dinger equation with $\omega_\alpha = \vect{k}_\alpha^2/(2m)$. The first-order perturbed wave is proportional to a Hankel function $H_0^{(1)}$ of the first kind:
\begin{alignat}{2}
    \Psi^{(1)} &= \sum_\alpha \Psi_\alpha \frac{-i G m}{4} e^{-i\omega_\alpha t} \\
    &\phantom{= \sum_\alpha} \times \int \dd^2 x' \, n_\psi(\vect{x}') H_0^{(1)}\left(|\vect{k}_\alpha| |\vect{x}-\vect{x}'|\right) e^{i \vect{k}_\alpha \cdot \vect{x}'}. \nonumber
\end{alignat}
Defining the (unperturbed) ``number density'' $n \equiv \sum_\alpha |\Psi_\alpha|^2$, the expected potentials to first- and second-order in $G$ are:
\begin{alignat}{2}
    V^{(0)}(\vect{x}) &= + \frac{G n}{2}; \\
    V^{(1)}(\vect{x}) &= - \frac{G^2 m n}{4} \int \dd^2 x' \, n_\psi(\vect{x}) \mathcal{F}_2(\vect{x}-\vect{x}').
\end{alignat}
The 2D form factor is:
\begin{alignat}{2}
    \hspace{-0.1em} \mathcal{F}_2(\vect{x}) 
    &= \mathrm{Re} \int \ddbar^2k \, f(\vect{k}) i H^{(1)}_0(|\vect{k}| |\vect{x}|)  e^{-i\vect{k}\cdot\vect{x}} \\
    &= \mathrm{Re} \int_0^\infty \frac{\dd k \, k}{\sigma_k^2} e^{-{k^2}/{2\sigma_k^2}} i H^{(1)}_0(k |\vect{x}|) J_0(k |\vect{x}|),
\end{alignat}
with the last line for a 2D Maxwell-Boltzmann distribution. It scales logarithmically at short distances, and falls off exponentially for $|\vect{x}| \gg \sigma_k^{-1}$, so that the $2+1$D (like the $3+1$D) wake force has a short range.

Perturbatively, the predicted number density can be written as $n(\vect{x}) \simeq  2(V^{(0)} + V^{(1)})/G$ to leading order in $G$. These perturbative predictions are overlaid as red solid lines on \Fig{fig:wake_sim_1D}, where they are found to match the numerical results well when the perturbative expansion parameter $\epsilon \equiv |V^{(1)} / V^{(0)}| = |G| m n_\psi / 2\sigma_k^2$, which is the same in $2+1$D and $3+1$D, is small in absolute value. The red bands in \Fig{fig:wake_sim_1D} are the corresponding fractional $\pm 1\sigma_\mathrm{th}$ theory errors around the first-order prediction with $\sigma_\mathrm{th} = \epsilon^2$. 
The permille-level concordance between the numerical results and the perturbative prediction indicates that the perturbative expansion parameter is $\epsilon$ in two or more spatial dimensions, even for large sources.

Finally, a simple thermodynamic argument can be used to estimate the density modulation deep inside a medium-dependent potential. Suppose particles outside the potential have a number density $n_\mathrm{outside} \propto \int \ddbar^d k \, e^{-\vect{k}^2/2\sigma_k^2} = \int \ddbar^d k \, e^{-E_\mathrm{kin}/T}$ given by a Maxwell-Boltzmann distribution with effective temperature $T = \sigma_k^2/m$. Modes deep inside a constant potential should have an effective Boltzmann factor $e^{-(E_\mathrm{kin} + G n_\psi/2)/T}$, thus yielding the nonperturbative prediction:
\begin{alignat}{2}
    \frac{n_\mathrm{inside}}{n_\mathrm{outside}} \simeq \exp \left\lbrace - \frac{G n_\psi/2}{T} \right\rbrace = \exp \left\lbrace - \frac{G m n_\psi}{2 \sigma_k^2} \right\rbrace, \label{eq:thermo}
\end{alignat}
assuming such equilibrium is indeed established. In the perturbative regime, the estimate agrees with that of the first-order wake potential: $n_\mathrm{inside} / n_\mathrm{outside} - 1 = V^{(1)}/V^{(0)} = - G m n_\psi / 2\sigma_k^2$, which holds both in $2+1$D and $3+1$D. The nonperturbative thermodynamic estimate from \Eq{eq:thermo} is indicated by the light blue dashed line in the right panel of \Fig{fig:wake_sim_1D}.
It yields a milder density decrease because the next-to-leading order in the perturbative expansion has opposite sign. 

\paragraph*{3+1 dimensions---}
\begin{figure}[htb!]
    \centering
    \includegraphics[width=0.45\textwidth]{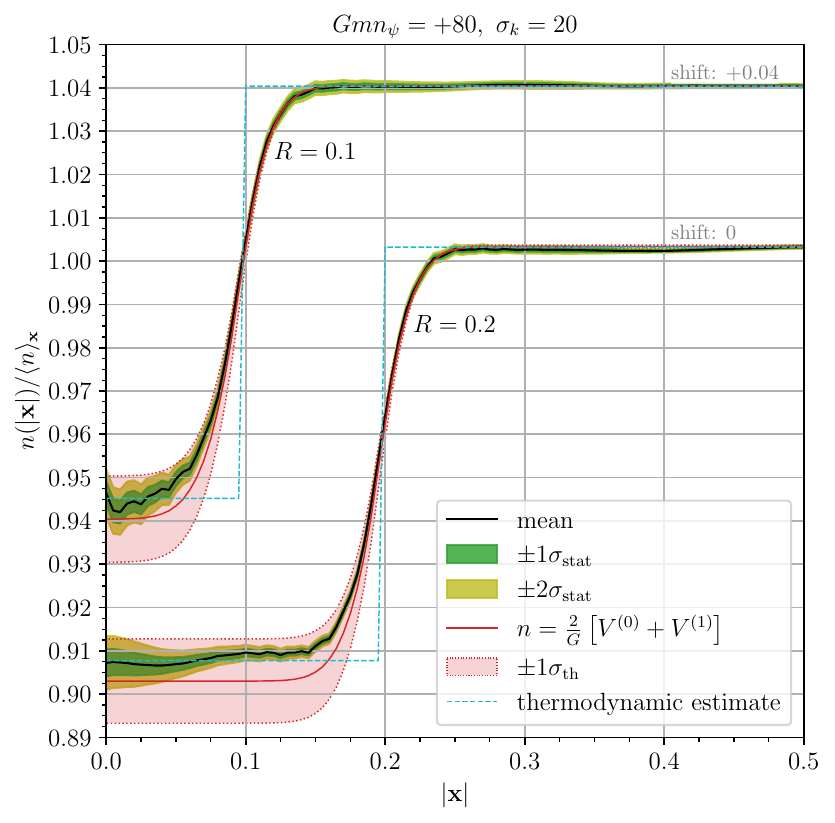}
    \caption{Number density $n(|\vect{x}|)$ relative to the 3-torus average $\langle n \rangle_{\vect{x}}$ in the $3+1$D simulations, averaged at a distance $|\vect{x}|$ from the center of a ball-shaped barrier with strength $G m n_\psi = +80$ and radius $R= \lbrace 0.1,0.2\rbrace$ (the former vertically offset by $+0.04$ for clarity), for an isotropic MB distribution with $\sigma_k = 20$.
    Labels are as in \Fig{fig:wake_sim_1D}.
    The estimated systematic error from partial field initialization inside the barrier is $\sigma_\mathrm{sys} \sim \epsilon (4\pi/3) R^3 \sim \lbrace 0.0004, 0.003 \rbrace$ for $|\vect{x}| < R$ but is not shown.}
    \label{fig:wake_sim_3D}
    \vspace{-1em}
\end{figure}
The results of the $3+1$D simulations are shown in \Fig{fig:wake_sim_3D} as black curves with green (yellow) statistical error bands at $\pm 1 \sigma_\mathrm{stat}$ ($\pm 2 \sigma_\mathrm{stat}$). The chosen numerical resolution is lower for computational feasibility reasons, but the qualitative features are similar to the $2+1$D case. The perturbative prediction for the first-order nonrelativistic wake potential was already derived in the main text (cfr.~\Eqs{eq:V_wake_scalar} and \ref{eq:form_MB}):
\begin{alignat}{2}
    V^{(1)}(\vect{x}) = - \frac{G^2 m n}{4\pi}  \int \dd^3 x' \, n_\psi(\vect{x}') \frac{e^{-2\sigma_k^2 (\vect{x}-\vect{x'})^2}}{|\vect{x}-\vect{x}'|}. \label{eq:wake_3D_app}
\end{alignat}
The corresponding perturbative predictions for the density of $\phi$ particles, $n(\vect{x}) = (2/G) ( G n / 2 + V^{(1)}(\vect{x}))$ are shown as red curves with $1\sigma_\mathrm{th} = \epsilon^2$ theory error bands. The thermodynamic estimate from \Eq{eq:thermo} is indicated by the light blue dashed line, and lines up precisely with the numerical results deep inside the barrier. The systematic uncertainty from partial field initialization inside the barrier is not shown, but should be of order $\sigma_\mathrm{sys} \sim \epsilon (4\pi/3) R^3$ for $|\vect{x}| < R$, and is negligible for $R = 0.1$ but more appreciable for $R = 0.2$, with the hierarchy $\sigma_\mathrm{stat} \lesssim \sigma_\mathrm{sys} \lesssim \sigma_\mathrm{th}$ in that case. Like in $2+1$D, the nonperturbative numerical results in $3+1$D are in quantitative agreement with perturbative wake potential calculations.

\section{Useful formulae} \label{app:formulae}
In this appendix, I collect some useful formulae for 3D wake potential calculations. 

Integrals to compute the inverse Fourier transform of matrix elements:
\begin{align}
&\int \ddbar^3q \, e^{i\vect{q}\cdot\vect{x}} \frac{1}{\vect{q}^2-\vect{k}^2-i\varepsilon} = \frac{e^{i|\vect{k}|r}}{4\pi r}, \label{eq:integral_1}\\
&\int \ddbar^3q \,  e^{i\vect{q}\cdot\vect{x}} \frac{\vect{q}}{\vect{q}^2-\vect{k}^2-i\varepsilon} = \frac{e^{i|\vect{k}|r}}{4\pi r} \hat{\vect{x}} \left[\frac{i}{r} - |\vect{k}|\right], \label{eq:integral_2} \\
&\int \ddbar^3q \, e^{i\vect{q}\cdot\vect{x}} \frac{(\vect{q}\cdot \vect{n_1})(\vect{q}\cdot \vect{n_2})}{\vect{q}^2-\vect{k}^2-i\varepsilon} \label{eq:integral_3} \\
&= \frac{e^{i|\vect{k}|r}}{4\pi r} \bigg\lbrace(\vect{n_1}\cdot\vect{n_2}) \left[-\frac{1}{r^2}+\frac{i|\vect{k}}{r}\right]\nonumber\\
&\phantom{= \frac{e^{i|\vect{k}|r}}{4\pi r} \bigg\lbrace}+(\hat{\vect{x}}\cdot\vect{n_1})(\hat{\vect{x}}\cdot\vect{n_2})\left[\frac{3}{r^2}-\frac{i|\vect{k}|}{r}-|\vect{k}|^2 \right]\bigg\rbrace . \nonumber
\end{align}
Exact and approximate ($q \ll m$) spinor identities:
\begin{alignat}{2}
    \slashed{k}u^s(k) &= m u^s(k), \label{eq:id_spinor_1} \\
    -\slashed{k}v^s(k) &= m v^s(k), \label{eq:id_spinor_2} \\
    \sum_s \bar{u}^s(k) \slashed{q} u^s(k) &= -4\vect{q}\cdot\vect{k} ,\label{eq:id_spinor_3} \\
    \sum_s \bar{v}^s(k) \slashed{q} v^s(k) &= -4\vect{q}\cdot\vect{k}; \label{eq:id_spinor_4} \\
    \bar{u}^{s'}(p-q) u^s(p) &\simeq 2 m \delta^{s's}, \label{eq:id_spinor_5}\\
    \bar{u}^{s'}(p-q) i \gamma^5 u^s(p) &\simeq i \vect{q} \cdot \vect{\sigma}_{s s'}, \label{eq:id_spinor_6} \\
    \bar{v}^{s'}(p-q) v^s(p) &\simeq - 2m \delta^{s' s} \label{eq:id_spinor_7}.
\end{alignat}
Nonrelativistic spinor identities:
\begin{alignat}{2}
    \bar{u}^{s'}(p) \gamma^\mu u^s(p) &\simeq 2 m \delta^{s's} \delta^{\mu 0} , \label{eq:id_spinor_8}\\
    \bar{u}^{s'}(p) \gamma^0 \gamma^5 u^s(p) &\simeq 0  \label{eq:id_spinor_9}, \\
    \bar{u}^{s'}(p) \vect{\gamma} \gamma^5 u^s(p) &\simeq 2 m \vect{\sigma}_{s's}.  \label{eq:id_spinor_10} 
\end{alignat}

\bibliography{wake}

\end{document}